\documentclass[useAMS,usenatbib]{mn2e}

\usepackage{amssymb}
\usepackage{amsmath}
\usepackage{times}
\usepackage{graphicx}

\newcommand{\za}{$z_{\rm abs}$}

\newcommand{\lya}{Ly$\alpha$}

\newcommand{\kms}{\mbox{km s$^{-1}$}}

\newcommand{\civ}{\hbox{C\,{\sc iv}}}

\newcommand{\Ociv}{\mbox{$\Omega_{\rm C\,\textsc{iv}}$}}
 
\newcommand{\apjs}{ApJS}
\newcommand{\apj}{ApJ}
\newcommand{\aj}{AJ}
\newcommand{\araa}{ARA\&A}
\newcommand{\apjl}{ApJL}
\newcommand{\nat}{Nature}

\newcommand{\aap}{A\&A}
\newcommand{\mnras}{MNRAS}

\def\ltsima{$\; \buildrel < \over \sim \;$}
\def\simlt{\lower.5ex\hbox{\ltsima}}
\def\gtsima{$\; \buildrel > \over \sim \;$}
\def\simgt{\lower.5ex\hbox{\gtsima}}

\title[Downturn in $\Omega_{\rm C IV}$ as $z=6$ is approached]
{A downturn in intergalactic C\,{\sc iv} as redshift 6 is approached}

\author[Ryan-Weber et al.]{Emma V. Ryan-Weber$^{1}$\thanks{ Current
    address: Centre for Astrophysics and Supercomputing, Swinburne
    University of Technology, PO Box 218, Victoria 3122,
    Australia. email: eryanweber@swin.edu.au}, Max Pettini$^1$, Piero
  Madau$^2$, and Berkeley J. Zych$^1$\\ $^1$Institute of Astronomy,
  Madingley Rd, Cambridge, CB3 0HA, UK\\ $^2$Department of Astronomy
  and Astrophysics, University of California, Santa Cruz, CA 95064,
  USA}

\begin{document}

\date{Accepted ... Received ... in original form ...}

\pagerange{\pageref{firstpage}--\pageref{lastpage}} \pubyear{2009}

\maketitle

\label{firstpage}

\begin{abstract}
We present the results of the largest survey to date for 
intergalactic metals at redshifts $z > 5$, using near-IR
spectra of nine QSOs with emission redshifts $z_{\rm em} > 5.7$.
We detect three strong \civ\ doublets at
$z_{\rm abs} = 5.7$--5.8, two low ionisation systems
at $z _{\rm abs} > 5$, and numerous Mg\,{\sc ii} absorbers
at $z_{\rm abs} = 2.5$--2.8. 
We find, for the first time, a change in the comoving mass density
of \civ\ ions as we look back to redshifts $z > 5$.
At a mean $\langle z \rangle = 5.76$, we deduce
$\Omega_{\rm C\,\textsc{iv}} = (4.4 \pm 2.6) \times 10^{-9}$
which implies a drop by 
a factor of $\sim 3.5$ compared to the value at 
$z < 4.7$, after accounting for the differing sensitivities
of different surveys.
The observed number of \civ\ doublets is also lower
by a similar factor, compared to expectations for a
non-evolving column density distribution of absorbers.

These results point to a rapid build-up of
intergalactic \civ\ over a period of only $\sim 300$\,Myr;
such a build-up could reflect the accumulation of metals 
associated with the rising levels of star formation 
activity from $z \sim 9$ indicated by galaxy counts, and/or
an increasing degree of ionisation of the intergalactic medium
(IGM), following 
the overlap of ionisation fronts from star-forming regions.
If the value of  $\Omega_{\rm C\,\textsc{iv}}$ we derive
is typical of the IGM at large, it would imply a metallicity
$Z_{\rm IGM} \simgt 10^{-4}\,Z_{\odot}$.
The early-type stars responsible for 
synthesising these metals would have emitted only 
about one Lyman continuum photon per baryon  
prior to $z = 5.8$; such a background is insufficient to 
keep the IGM ionised
and we speculate on possible factors which could make up the
required shortfall.
\end{abstract}


\section{Introduction}

It is quite remarkable that elements heavier than
helium exist in the intergalactic medium
(IGM) only $\sim 1$\,Gyr after the Big Bang,
as indicated by the presence of metal absorption
lines associated with the \lya\ forest of high redshift 
QSOs (Songaila 2001; Becker et al. 2006; 
Ryan-Weber, Pettini \& Madau 2006; Simcoe 2006).
For these elements to be observed there
needs to be significant star formation prior to redshift $z = 6$, 
and the nucleosynthetic products of such early stars need to be 
expelled beyond the gravitational influence of their host galaxies.
The production of early metals is closely linked to the
emission of the Lyman continuum (LyC) photons responsible for ionizing
the IGM and ending the so-called ``dark ages'' 
(Gnedin \& Ostriker 1997; Fan, Carilli, \& Keating 2006),
a process which could have started as early as $z =  20$
and continued until completion at $z \sim  6$ (Dunkley et al. 2009).

Clearly, some galaxies were already actively forming stars at $z > 6$.
The current record for a spectroscopically confirmed galaxy is
held by the \lya\ emitter at $z = 6.96$ discovered by Iye et al. (2006),
but several candidates selected via the redshifted Lyman break
have been found at $z > 7$ (Stark et al. 2007; Richard et al. 2008;
Bouwens et al. 2008 and references therein).
Furthermore, analyses of the broad-band spectra of galaxies
at redshifts $z = 2 - 6$ have in several cases revealed objects
with `mature' stellar populations, suggesting that star
formation in these galaxies was underway well before $z = 6$
(Shapley et al. 2001; Shapley et al. 2005; Yan et al. 2006; Eyles et al. 2007).

Whether the level of star formation so far accounted for at
$z > 6$ is in fact sufficient to maintain reionization is far less
certain, because it depends on a number
of essentially unknown parameters, especially
the escape fraction of Lyman continuum photons from
the sites of star formation, the degree of clumping of the sources
and of the IGM,
and the slope of the faint end of the galaxy luminosity function
at these early times (Madau, Haardt, \& Rees 1999).
Using the luminosity function of Bouwens et al. (2006) at $z \simeq 6$
and assuming an escape fraction of 20\%,  
Bolton \& Haehnelt (2007) concluded that 
there may be \textit{just enough} LyC photons 
to keep the IGM ionised, suggesting that 
cosmic reionization was a `photon
starved' process (see also Gnedin 2008).
QSOs are unlikely to help matters, 
since their contribution to the metagalactic
photoionization rate at these redshifts is estimated
to be at most 15\% (Srbinovsky \& Wyithe 2007).
Bolton \& Haehnelt (2007) proposed that, in order to
complete reionization by $z \simeq 6$, either
the escape fraction of LyC photons must have been higher
at earlier times, or the ionizing spectrum must have been harder (or both).
Alternatively, the faint end slope of the luminosity function 
may have been steeper at $z > 6$ compared to the
luminosity function determined by Bouwens et al. (2006),
as indeed proposed by Stark et al. (2007).

The cosmic budget of elements produced via stellar nucleosynthesis
(loosely referred to as metals) provides an entirely different,
and complementary, census of the star-formation activity prior
to a given epoch. Indeed, at redshifts $z = 2 - 3$ there appears to
be an approximate agreement (within a factor of about two)
between the observed metal budget---summing together the
contributions of stars, IGM and diffuse gas in galaxies---and 
expectations based on the integral of the cosmic star formation rate
density over the first 2-3\,Gyr of the Universe history
(Pettini 2006; Bouch{\'e} et al. 2007).

At higher redshifts, we rely increasingly on the \lya\ forest
and gamma-ray burst afterglows to trace metals in the IGM
and in galaxies, since 
at $z > 5$ the spectra of even the most luminous galaxies are 
too faint to be studied in detail with current instrumentation.
The most commonly encountered metal lines associated
with the \lya\ forest are the 
C\,{\sc iv}~$\lambda\lambda 1548.2041, 1550.7812$ doublet
(Cowie et al. 1995; Ellison et al. 2000). 
Photoionisation modelling 
(Schaye et al. 2003; Simcoe, Sargent, \& Rauch 2004)
suggests that at redshifts $z = 2 - 3$ the carbon
abundance is a function of the gas density,
as expected (Cen \& Ostriker 1999),
and that the mean metallicity of the IGM
is $Z_{\rm IGM} \approx 1/1000\,Z_{\odot}$.

Interestingly, the comoving mass density of triply
ionised carbon, obtained by integrating the distribution
of column densities $N$(C\,{\sc iv}), appears to be
approximately constant from $z \simeq 5$ to $z \simeq 1.5$:
expressed as a fraction of the 
critical density, 
$\Omega_{\rm C\,\textsc{iv}} \simeq (1 - 4) \times 10^{-8}$
(Songaila 2001; Pettini et al. 2003\footnote{The values reported by those authors
have been updated to the same
$H_0=70$\,\kms\,Mpc$^{-1}$, $\Omega_{\rm M} = 0.3$ and
$\Omega_{\Lambda} = 0.7$ cosmology used in this paper.})
over a period of $\sim 3$\,Gyr which 
saw the peak of the cosmic star-formation and massive
black hole activity (e.g. Reddy et al. 2008 and references therein).
This surprising result has led to conflicting interpretations
as to the source of these intergalactic metals, with some
authors (e.g. Madau, Ferrara, \& Rees 2001; Porciani \& Madau 2005)
proposing an origin in low-mass galaxies at $z \gg 6$,
when the pollution of large volumes of the IGM would have
been easier, while others (e.g Adelberger 2005) place them
much closer to the massive galaxies responsible for most of
the metal production at redshifts $z = 1.5$--5.
While both processes presumably contribute, 
their relative importance has yet to be established
with certainty (Songaila 2006).

The most extensive theoretical work in this area has been
published by B.~D.~ Oppenheimer  and collaborators. 
In a series of papers (Oppenheimer \& Dav{\'e} 2006, 2008;
Dav{\'e} \& Oppenheimer 2007) these authors analysed the 
output of cosmological simulations of galaxy formation
which include feedback in the form of momentum-driven winds.
They found that the apparent lack of evolution of
$\Omega_{\rm C\,\textsc{iv}}$ from $z = 5$ to 1.5 can be explained
as the result of two countervailing effects: an overall increase
of the cosmic abundance of carbon, reflecting the on-going
pace of star formation, and an accompanying reduction in the 
fraction of carbon which is triply ionised.

Whether this is the true picture or not, it is clear that determining
$\Omega_{\rm C\,\textsc{iv}}$ at $z > 5$ is the next observational
priority; in any case, metal lines are our only remaining probe
of the IGM at these high redshifts where the \lya\ forest itself
becomes effectively opaque. 
Such an extension of metal-line surveys
in QSO spectra necessitates observations at moderate spectral
resolution and signal-to-noise ratio at near-infrared (near-IR)
wavelengths, a regime which has only recently begun to be exploited
for such purposes (e.g. Kobayashi et al. 2002; Nissen et al. 2007).
In a pilot study (Ryan-Weber, Pettini, \& Madau  2006),
we demonstrated that QSO  absorption line spectroscopy 
could indeed be performed in the near-IR to the levels required to
detect intergalactic C\,{\sc iv} absorption. 
From observations of two QSO at redshifts 
$z_{\rm em} = 6.28$ and 5.99 we discovered two
strong C\,{\sc iv} doublets at absorption redshifts
$z_{\rm abs} = 5.7238$ and 5.8290. 
These two absorbers, which were 
confirmed by an independent study by Simcoe (2006), 
would---if typical--- imply that a
high concentration of C\,{\sc iv} was already in place
in the IGM at $z \simeq 6$, but clearly much better
statistics than  those afforded by only two QSO sight-lines
are necessary to assess the true level of  $\Omega_{\rm C\,\textsc{iv}}$.

Given the success of our pilot study, in the last two years
we have conducted an observational campaign on the Keck
and ESO-VLT telescopes aimed at securing near-IR spectra
of as many as possible of the 19 known QSOs with
$z_{\rm em} > 5.7$ and infrared magnitude $J<19.2$.
The results of this programme are presented here.
Specifically, we have observed 13 QSOs and obtained
useful data for ten of them. This sample represents
an increase by a factor of seven in the pathlength
probed, from the absorption distance
$\Delta X=3.6$ covered by Ryan-Weber et al. (2006) to
$\Delta X=25.1$.\footnote{The absorption
distance $X$ is defined so that
$dX/dz = H_0(1 + z)^2/H(z)$, where $H$ is the Hubble
parameter. Populations of absorbers
with constant physical cross sections and comoving number
densities maintain constant line densities $dN/dX$ as they passively
evolve with redshift.}
With this extended data set, we find the first 
evidence for a \textit{decrease} in $\Omega_{\rm C\,\textsc{iv}}$
from $z \sim 4.7$ to $z \sim 5.8$, in broad agreement
with the predictions of the cosmological simulations
by Oppenheimer, Dav{\'e}, \& Finlator (2009).

This paper is organised as follows.
Section~2 describes the observations, followed by 
brief comments on each QSO in Section~3.
In Section 4 we deduce the value of \Ociv\
implied from our three detections of \civ\
absorbers within the redshift range $5.2 \leq z_{\rm abs} \leq 6.2$.
In Sections 5 and 6, we consider the completeness of the survey,
and discuss the evidence for a drop in \Ociv\ as we look
back to redshifts $z > 4.7$.
The implications of these results for early star formation and
reionisation are discussed in Section~7. 
We finally summarise our
main conclusions in Section~8.

Throughout this paper we use a 
`737' cosmology with $H_0=70$\,\kms\,Mpc$^{-1}$, 
$\Omega_{\rm M} = 0.3$ and
$\Omega_{\Lambda} = 0.7$. Much of the previous
work (including our own) 
with which we compare the present results 
was based on an Einstein-de Sitter cosmology; in all cases, we
have converted the quantities of interest to the cosmology adopted
here.

\begin{table*}
\centering
\begin{minipage}[c]{1.2\textwidth}
    \caption{\textsc{Details of the Observations}}
    \begin{tabular}{@{}lclcrcrrc}
    \hline
    \hline
   \multicolumn{1}{c}{QSO}
& \multicolumn{1}{c}{$J_{\rm Vega}$} 
& \multicolumn{1}{c}{$z_{\rm em}$}
& \multicolumn{1}{c}{Telescope/}
& \multicolumn{1}{c}{Wavelength}
& \multicolumn{1}{c}{Resolution}
& \multicolumn{1}{c}{Integration}
& \multicolumn{1}{c}{S/N$^{\rm a}$}
& \multicolumn{1}{c}{$\Delta z^{\rm b}$}\\
   \multicolumn{1}{c}{ }
& \multicolumn{1}{c}{(mag)}
& \multicolumn{1}{c}{ }
& \multicolumn{1}{c}{Instrument }
& \multicolumn{1}{c}{ Range (\AA)}
& \multicolumn{1}{c}{(km~s$^{-1}$)}
& \multicolumn{1}{c}{Time (s)}
& \multicolumn{1}{c}{ }
& \multicolumn{1}{c}{}\\
    \hline
ULAS\,J020332.38+001229.2            & 19.1    & 5.706$^{\rm c}$         & Keck\,{\sc ii}/NIRSPEC  & 10\,219--10\,588~           & 185      &  6\,300~~~~          & 10--13   & 5.601--5.639$^{\rm d}$ \\
SDSS\,J081827.40+172251.8             & 18.5     & 6.00               & VLT1/ISAAC              & 9\,877--10\,790$^{\rm e}$   & ~\,53    & 32\,400$^{\rm f}$~~  &  8--13   & 5.380--5.930$^{\rm e}$ \\
SDSS\,J083643.85+005453.3             & 17.9     & 5.810$^{\rm g}$    & VLT1/ISAAC              & 10\,100--10\,550~           & ~\,53    & 16\,200~~~~          &  8       & 5.524--5.752~ \\
SDSS\,J084035.09+562419.9             & 19.0     & 5.774$^{\rm c}$    & Keck\,{\sc ii}/NIRSPEC  & 9\,548--11\,140~            & 185      & 11\,700~~~~          & 10--22   & 5.167--5.706~ \\
SDSS\,J103027.01+052455.0$^{\rm h}$   & 18.9 & 6.309$^{\rm i}$    & VLT1/ISAAC              & 9\,880--10\,780$^{\rm e}$   & ~\,53    & 82\,800$^{\rm j}$~~~ &  5--7    & 5.382--5.951$^{\rm e}$ \\
SDSS\,J113717.73+354956.9             & 18.4     & 5.962$^{\rm c}$    & Keck\,{\sc ii}/NIRSPEC  & 9\,542--11\,172~            & 185      &  9\,000~~~~          & 20--40   & 5.163--5.893~ \\
SDSS\,J114816.64+525150.2             & 18.1     & 6.421$^{\rm k}$     & Keck\,{\sc ii}/NIRSPEC  & 9\,572--11\,159~            & 185      & 18\,000~~~~          & 16--30   & 5.183--6.196~ \\
SDSS\,J130608.26+035626.3$^{\rm h}$   & 18.8 & 6.016$^{\rm i}$    & VLT1/ISAAC              & 9\,914--10\,332~            & ~\,53    &  7\,200~~~~          &  5       & 5.404--5.662~ \\
SDSS\,J160253.98+422824.9             & 18.5     & 6.051$^{\rm c}$    & Keck\,{\sc ii}/NIRSPEC  & 9\,553--11\,143~            & 185      &  9\,900~~~~          & 17--48   & 5.170--5.981~ \\
SDSS\,J205406.49$-$000514.8           & 19.2    & 6.062              & Keck\,{\sc ii}/NIRSPEC  & 9\,681--10\,927~            & 185      &  5\,400~~~~          &  8--12   & 5.253--5.991~ \\
   \hline
    \end{tabular}
    \smallskip

$^{\rm a}$Signal-to-noise ratio over wavelength range sampled.\\
$^{\rm b}$Redshift range covered for C\,{\sc iv} absorption.\\
$^{\rm c}$When available, we quote values of $z_{\rm em}$ measured 
from the C\,{\sc iv}~$\lambda 1549.062$ broad emission line in our NIRSPEC spectra.
As is well known, \\
the QSO systemic redshift, $z_{\rm sys}$, can be higher than the value measured from
high ionisation lines such as C\,{\sc iv}, by up 
to several thousand km~s$^{-1}$ \\
(e.g. Richards et al. 2002).\\
$^{\rm d}$BAL QSO; not included in the survey statistics.\\
$^{\rm e}$With a small gap between two wavelength settings.\\
$^{\rm f}$Two wavelength settings, with exposure times of 14\,400 and 18\,000\,s respectively.\\
$^{\rm g}$Kurk et al. (2007).\\
$^{\rm h}$Observations reported in Ryan-Weber et al. (2006).\\
$^{\rm i}$Jiang et al. (2007).\\ 
$^{\rm j}$In three, partially overlapping, wavelength settings, with exposure times of 39\,600, 7\,200 and 36\,000\,s respectively.\\
$^{\rm k}$$z_{\rm em}$ measured from the Si\,{\sc iv}~$\lambda 1396.752$ broad emission line in our NIRSPEC spectrum.\\

     \label{tab:obs}
\end{minipage}
\end{table*}

\section{Observations and data reduction}

A trawl of the relevant literature reveals that 19
QSOs are currently known with emission redshift
$z_{\rm em} > 5.7$ and $J$ magnitude brighter than  $J \simeq 19.2$.
This magnitude is at the faint limit for recording medium-resolution 
near-IR spectra of S/N\,$\simgt 5$ with integration times of 
a few hours on 8-10\,m class telescopes using available
instrumentation. The redshift cut is required to give a useful
pathlength over which to search for C\,{\sc iv} absorption
at redshifts $z_{\rm abs} > 5$, this being the limit of previous surveys
(Songaila 2001; Pettini et al. 2003).

Over the four-year period between 2004 and 2008 we observed
13 of these 19 possible targets with NIRSPEC on the Keck\,{\sc ii}
telescope (McLean et al. 1998) and with ISAAC on VLT-UT1 (Moorwood et al. 1998).
For ten QSOs we secured spectra with S/N\,$\geq 5$;
relevant details are collected in Table~\ref{tab:obs}.
In the other three cases, the object was not pursued
after initial test exposures, either because the spectrum
displayed broad absorption lines 
(SDSS\,J104433.04$-$012502.2 at $z_{\rm em} = 5.74$
and SDSS\,J104845.05$+$463718.3, $z_{\rm em} = 6.20$), 
or because the QSO turned out to be fainter than anticipated 
(
SDSS\,J141111.29$+$121737.4, $z_{\rm em} = 5.93$).
The values of emission redshift $z_{\rm em}$ listed in Table~\ref{tab:obs}
are those measured from the discovery spectra, unless otherwise noted.

\subsection{NIRSPEC Observations}

With NIRSPEC we used the NIRSPEC~1$+$thin blocker filter
combination to record the wavelength region $\sim 9540$--11\,170\,\AA\
on the ALADDIN-3 InSb detector which has 
$1024\times1024$ 27$\mu$m pixels. 
With the $42\times 0.57$\,arcsec entrance slit, the 
spectrograph delivers a resolution 
${\rm FWHM} = 6.4$\,\AA\ (185\,km~s$^{-1}$
at 10\,355\,\AA), sampled with 2.1 pixels.
Typically, we exposed on an object for 900\,s
before reading out the detector, moving
the target by a few arcseconds along the slit,
and starting the subsequent exposure.
Total integration times ranged between 
5400 and 18\,000\,s (see column (7) of Table~\ref{tab:obs}).
All of the QSO observed were sufficiently bright
at near-IR wavelengths to be visible on the slit,
thereby facilitating guiding.


\begin{figure*}
\vspace{-2cm}
\centerline{\hspace{0.5cm}
\includegraphics[width=1.85\columnwidth,clip,angle=0]{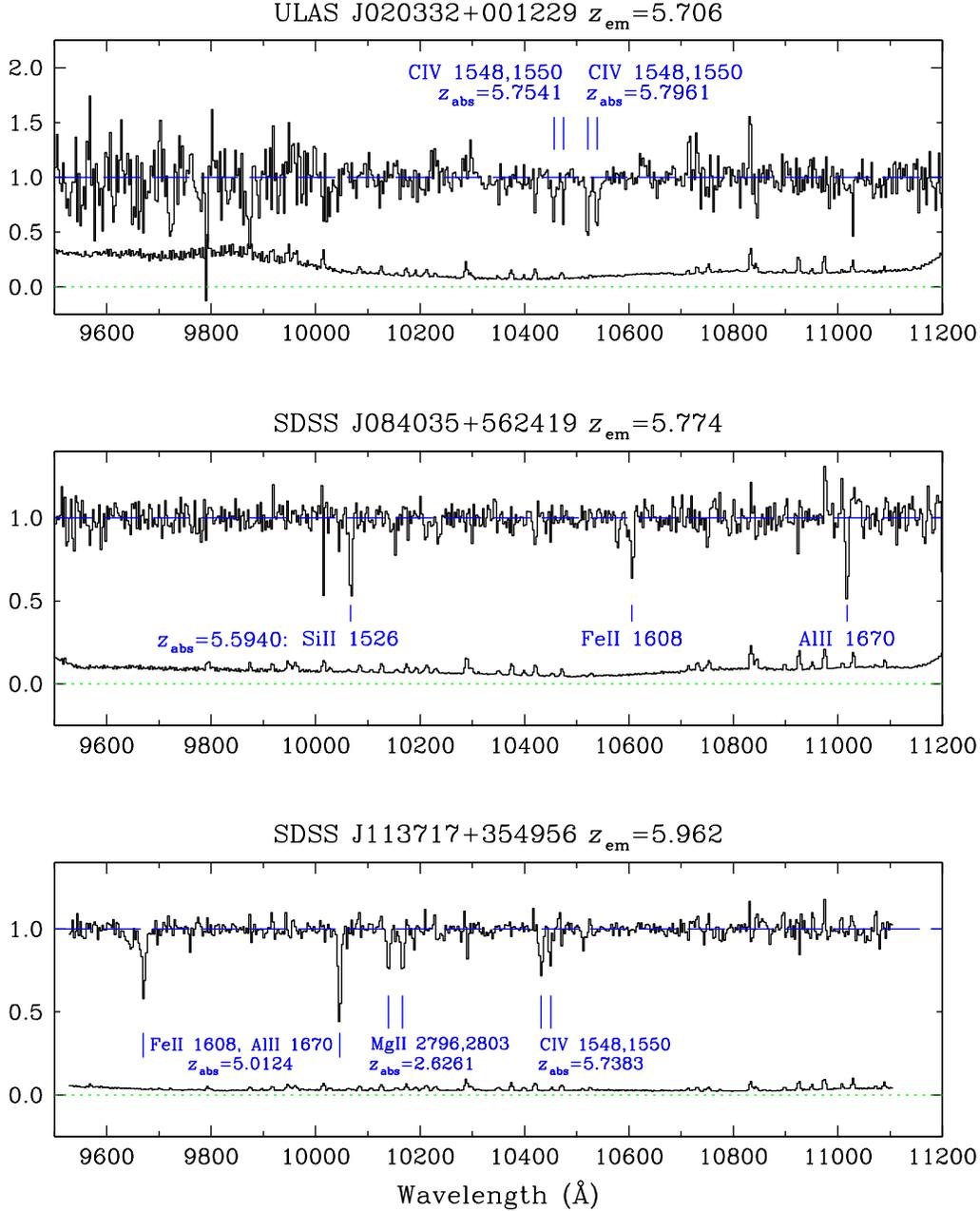}}
\vspace{-2cm}
\caption{NIRSPEC spectra of ULAS\,J020332.38+001229.2,
  SDSS\,J084035.09+562419.9 and SDSS\,J113717.73+354956.9. 
  In each panel, the top
  histogram shows the normalised QSO spectrum, while the lower
  histogram is the corresponding $1 \sigma$ error spectrum.
  Absorption features are indicated by vertical tick marks and
  labelled with their identification. 
  The two \civ\ absorption doublets seen in
  the spectrum of ULAS\,J020332.38+001229.2 (top panel)
  are close to the QSO emission redshift and are probably
  intrinsic (see section~\ref{sec:QSO1}).
  Parameters of the \civ\ doublet identified towards 
  SDSS\,J113717.73+354956.9 (bottom panel) are listed in
  Tables~\ref{tab:ew} and \ref{tab:abs}.}
\label{fig:spec1}
\end{figure*}

The NIRSPEC two-dimensional (2-D) spectra were
processed with a suite of purpose-designed
IDL routines kindly provided by Dr. G. Becker.
As described in more detail by Becker, Rauch, \& Sargent
(2009), this software does a very good job at 
subtracting the dark current, removing detector defects,
flat-fielding, and particularly removing the sky background
(achieved by employing the optimal sky subtraction
techniques of Kelson 2003)
which can be the limiting factor to the S/N achievable at
near-IR wavelengths.
Optimal extraction (Horne 1986)
is also used to trace the QSO signal in the 
processed 2-D images and
output a vacuum-heliocentric wavelength calibrated
one-dimensional (1-D) source spectrum and associated error.
The wavelength calibration uses the OH sky emission lines 
superposed on the QSO spectrum, with the
wavelengths tabulated in the atlas by Rousselot et al. (2000).

As we have many ($\geq 6$) individual 900\,s spectra 
(and associated error) of each of our QSOs, 
we combined them using a weighted mean algorithm
that operates on a pixel-by-pixel basis and rejects 
outliers beyond a specified number of standard deviations
from the mean (usually $\geq 3 \sigma$).
The spectra were finally normalised by dividing by
a spline fit to the underlying QSO continuum.
The rms deviation of the data from this continuum 
(away from obvious spectral features) provides an empirical
estimate of the final S/N achieved.\footnote{Strictly speaking,
this is a \textit{lower limit} to the real S/N of the spectra,
since it assumes a `perfect' fit to the QSO continuum. However,
in practice this is the value to be used in assessing the significance 
of any absorption features present in the spectrum.} 
We found this empirical estimate of the S/N to be lower,
by factors between 2.0 and 1.1, than
the pixel-to-pixel error on the weighted mean calculated
when co-adding the individual exposures, as described above.
This difference presumably reflects the residual importance
of systematic (rather than random) errors; in any case,
the values listed in the penultimate column of Table~\ref{tab:obs}
are those measured from the residuals about the continuum fit.
Since the S/N varies along each spectrum, we give in the Table
the maximum and minimum values which apply to the wavelength region
searched for absorption lines (listed for each QSO in column (5) of
Table~\ref{tab:obs}).


\begin{figure*}
\vspace{-2cm}
\centerline{\hspace{0.5cm}
\includegraphics[width=1.85\columnwidth,clip,angle=0]{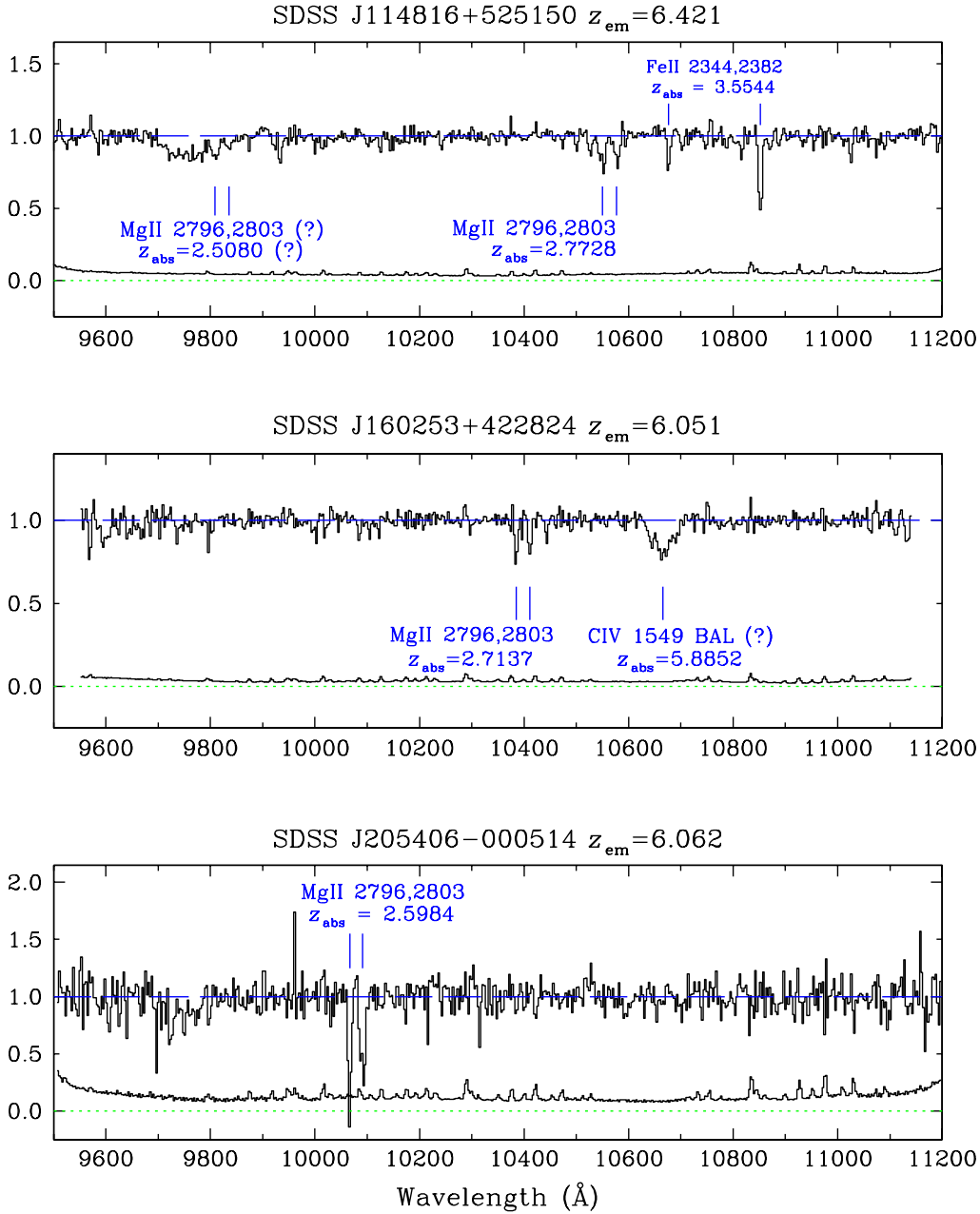}}
\vspace{-2cm}
\caption{NIRSPEC spectra of SDSS\,J114816.64+525150.2,
  SDSS\,J160253.98+422824.9 and SDSS\,J205406.49$-$000514.8. 
  In each panel, the top
  histogram shows the normalised QSO spectrum, while the lower
  histogram is the corresponding $1 \sigma$ error spectrum.
  Absorption features are indicated by vertical tick marks and
  labelled with their identification.}
\label{fig:spec2}
\end{figure*}

\subsection{ISAAC Observations}

The ISAAC spectra were obtained with a mixture
of service-mode and visitor-mode observations 
We used the instrument in its
medium resolution mode with a 0.6\,arcsec wide
slit, which results in a spectral resolution 
${\rm FWHM} = 1.8$\,\AA\ (53\,km~s$^{-1}$),
sampled with four pixels of the $1024\times1024$ Hawaii Rockwell 
array. With this set-up, the wavelength range spanned
is $\Delta \lambda \simeq 460$\, \AA; 
we normally chose a grating setting which placed the 
C\,{\sc iv} emission line at the red limit of the range covered.
In some cases, multiple wavelength settings were used, to increase
the pathlength for absorption covered (see Table~\ref{tab:obs}).
The observations were conducted in the 
conventional `beam-switching' mode, in a series of $4\times 900$\,s
exposures at two locations on the detector, typically separated 
by 10\,arcsec. After each cycle of four integrations, 
the object was moved along the slit to a new pair of
positions. The instrument was aligned on the sky at an angle chosen
to include on the slit a bright star which provides a useful pointing
and seeing reference.

The reduction of the ISAAC data used mainly IRAF tasks
following the steps
already described by Ryan-Weber et al. (2006). 
Briefly, we started from the ESO pipeline products 
which are 2-D images that have undergone
a substantial amount of pre-processing, including flat-fielding,
background subtraction, and wavelength calibration.
Given the weakness of the sources at the moderately high
dispersion of the ISAAC spectra, we found it most effective
to combine all the processed frames of a given QSO 
(at a given wavelength setting) in 2-D, after appropriate
shifts to bring them into alignment. This co-addition also
allowed us to reject deviant pixels.

One-dimensional spectra were extracted from the co-added
2-D frames, with a second pass at subtracting any residual sky features;
the data were then rebinned to 
two wavelength bins per resolution element 
in order to improve the S/N while still
maintaining adequate spectral sampling.
The final steps, including normalisation to the QSO continuum
and estimate of the rms noise, were the same as described for the NIRSPEC spectra.
As can be appreciated from inspection of Table~\ref{tab:obs},
the values of S/N achieved with ISAAC tend to be lower than those
of the NIRSPEC spectra, despite the generally longer integration times.
However, the significantly higher spectral resolution provided by ISAAC,
by a factor of $\sim 3.5$, largely compensates for the lower S/N,
so that the two sets of data reach comparable sensitivities in terms of
the minimum absorption line equivalent widths detectable.


\begin{figure*}
\vspace{-2cm}
\centerline{\hspace{0.5cm}
\includegraphics[width=1.85\columnwidth,clip,angle=0]{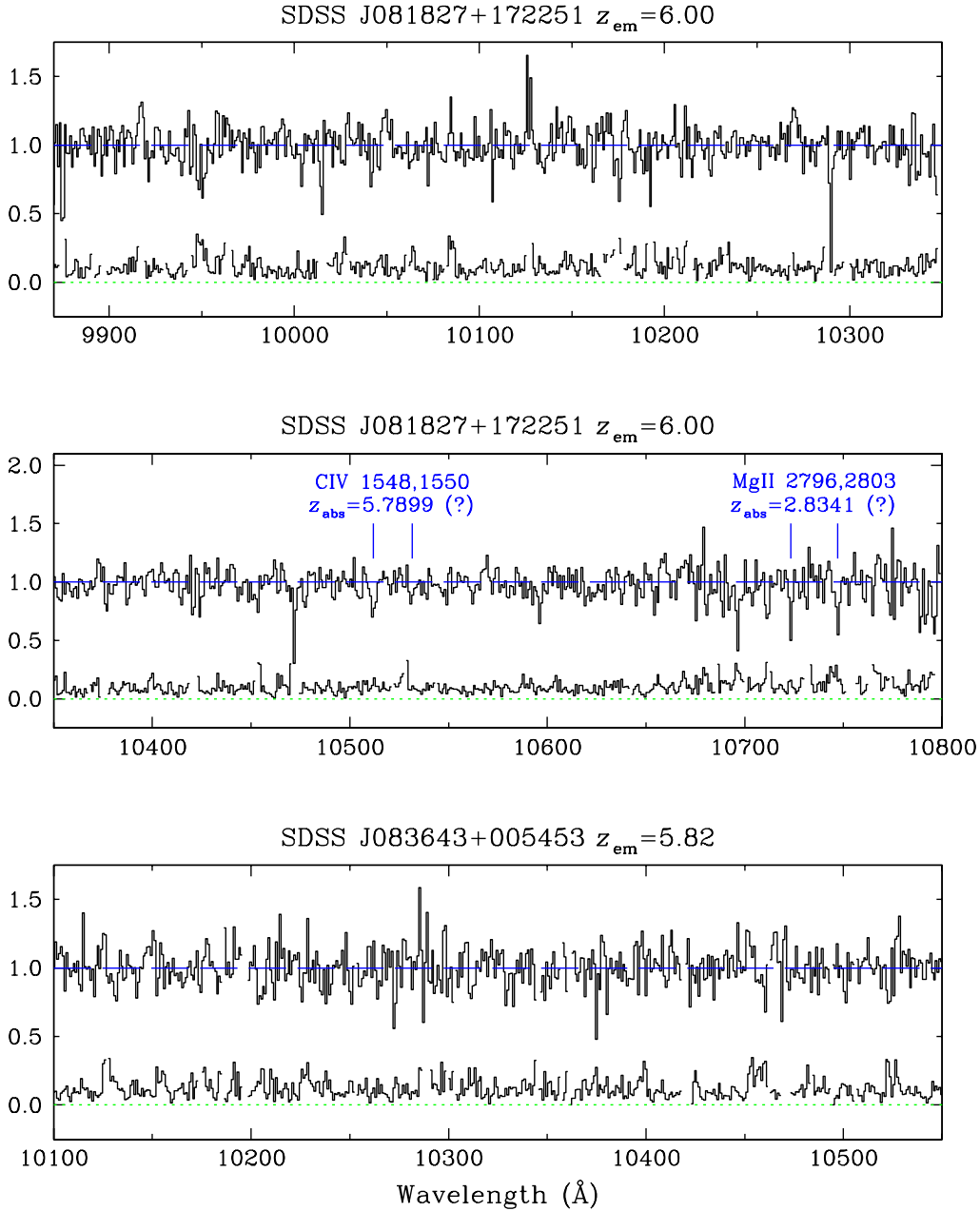}}
\vspace{-2cm}
\caption{ISAAC spectra of SDSS\,J081827.40+172251.8 (two wavelength
  settings) and SDSS\,J083643.85+005453.3.
  In each panel,  the top histogram shows the
  normalised QSO spectrum, while the lower histogram is the
  corresponding $1 \sigma$ error spectrum.  Absorption features are
  indicated by vertical tick marks and labelled with their
  identification. 
  There is a tentative detection of an 
  intervening \civ\ absorber towards
  SDSS\,J081827.40+172251.8 at $z_{\rm abs}=5.7899$;
  this possible system was omitted from the
  sample of definite detections used to estimate \Ociv.}
\label{fig:spec3}
\end{figure*}


\begin{figure*}
\vspace{-2cm}
\centerline{\hspace{0.5cm}
\includegraphics[width=1.85\columnwidth,clip,angle=0]{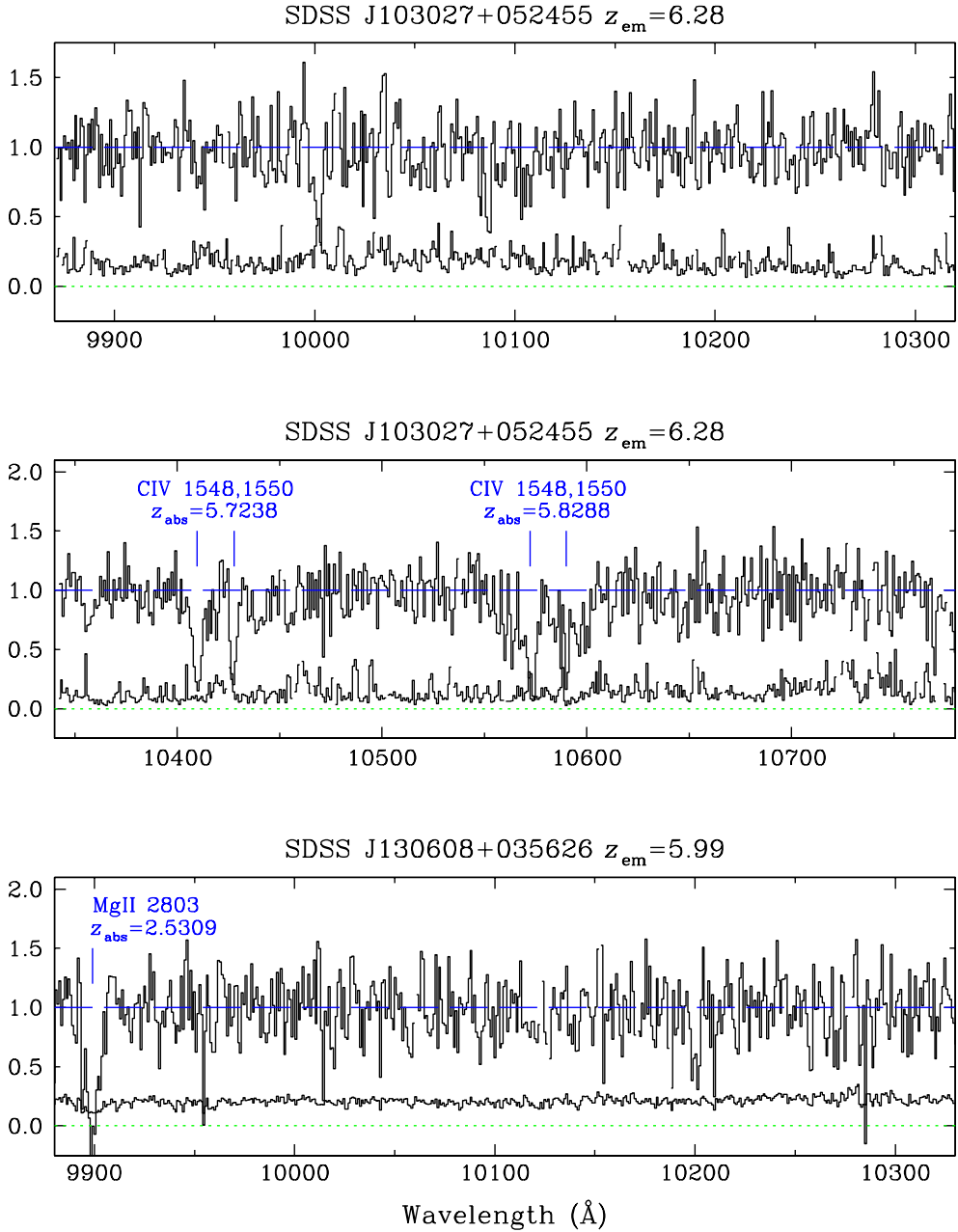}}
\vspace{-2cm}
\caption{ISAAC spectra of SDSS\,J103027.01+052455.0 (two wavelength
  settings) and SDSS\,J130608.26+035626.3. 
  In each panel, the top histogram shows the
  normalised QSO spectrum, while the lower histogram is the
  corresponding $1\sigma$ error spectrum.  
  Absorption features are
  indicated by vertical tick marks and labelled with their
  identification. Parameters of
  the two \civ\ doublets identified towards
  SDSS\,J103027.01+052455.0 are listed in Tables~\ref{tab:ew} and \ref{tab:abs}.}
\label{fig:spec4}
\end{figure*}

\section{Notes on individual QSOs}
\label{sec:QSOs}

The normalised spectra of the ten QSOs in Table~\ref{tab:obs} 
are reproduced in Figures~\ref{fig:spec1}--\ref{fig:spec4}, together 
with the corresponding error spectra.  
Absorption lines were identified
from a visual inspection of the spectra; 
they are labelled in Figures~\ref{fig:spec1}--\ref{fig:spec4}
and tabulated in Table~\ref{tab:ew}.
We now briefly comment on each QSO spectrum in turn.

\subsection{ULAS\,J020332.38+001229.2}
\label{sec:QSO1}

This QSO was identified by Venemans et al. (2007)
from the first data release of the UKIRT Infrared Deep Sky Survey 
(UKIDSS), and independently by Jiang et al. (2008) 
from the \textit{Sloan Digital Sky Survey} (SDSS).
Both surveys reported similar emission
redshifts: $z_{\rm em} = 5.86$ (Venemans et al.)
and $z_{\rm em} = 5.854$ (Jiang et al.).
From our NIRSPEC spectrum, however, we measure the lower
value $z_{\rm em} = 5.706$ from a broad emission feature
which we presume to be C\,{\sc iv}\,$\lambda 1549.062$;
this value is in much better agreement with the recent 
reassessment of the optical and infrared spectrum 
of this source by Mortlock et al. (2008) who conclude that
$z_{\rm em} = 5.72$. 
Mortlock et al. point out that this is a 
broad absorption line (BAL) QSO which led
the discoverers to misidentify the 
N\,{\sc v}\,$\lambda 1240.1$ emission line as \lya: 
the latter is completely absorbed by the 
blueshifted N\,{\sc v} absorption trough.

As this is a BAL QSO, we do not include it in the survey
for \civ\ absorption (as we are primarily interested in 
C\,{\sc iv} in the intergalactic medium, and wish to 
avoid contamination of our sample by metal-enriched gas
which may be ejected from the QSO). 
Nevertheless, it is of interest to note 
that we do detect two narrow \civ\ doublets at redshifts
$z_{\rm abs} = 5.7541$ and $z_{\rm abs} = 5.7961$
(see Figure~\ref{fig:spec1});
their equivalent widths are given in Table~\ref{tab:ew}. 
Relative to $z_{\rm em} = 5.706$, these two absorbers
have velocities $v = +2150$ and $+4025$\,km~s$^{-1}$
and are thus most likely associated with the QSO environment.
Positive velocities relative to $z_{\rm em}$ by up
to $\sim 2000$--3000\,km~s$^{-1}$ have been encountered
before among high ionisation absorbers at lower redshifts
(e.g. Fox, Petitjean, \& Bergeron 2008; Wild et al. 2008);
the values we deduce here are rather extreme and may in
part reflect the uncertainty in the systemic redshift
$z_{\rm sys}$.

\subsection{SDSS\,J081827.40+172251.8}
\label{sec:QSO2}

This $z_{\rm em} = 6.00$ QSO was observed with ISAAC at
two wavelength settings which, together, cover the wavelength
region 9877--10\,790\,\AA\ with only a very small gap between them
(see top two panels of Figure~\ref{fig:spec3}).
We find two possible absorption systems: a \civ\ doublet
at $z_{\rm abs} = 5.7899$ and a 
Mg\,{\sc ii}\,$\lambda\lambda 2796.3543, 2803.5315$ 
doublet at $z_{\rm abs} = 2.8341$; their reality 
can only be confirmed with higher S/N data.


 \begin{table*}
 \begin{minipage}{135mm}
 \caption{\textsc{Absorption Lines Identified}}
   \begin{tabular}{lllllll} 
\hline 
\hline
QSO & $z_{\rm em}$ & $z_{\rm abs}$ & Ion & $\lambda_{\rm lab}^{a}$(\AA) & $W_0$  (\AA) & Comments \\ 
\hline 
ULAS\,J020332.38+001229.2   & 5.706  & 5.7540 & C\,{\sc iv} & 1548.2041  & 0.38 & Intrinsic\\
                                               &           & 5.7543 & C\,{\sc iv} & 1550.7812  & 0.25 & Intrinsic\\
                                               &           & 5.7959 & C\,{\sc iv} & 1548.2041  & 0.62 & Intrinsic\\
                                               &           & 5.7962 & C\,{\sc iv} & 1550.7812  & 0.56 & Intrinsic\\
\\
SDSS\,J081827.40+172251.8     & 6.00   & 5.7897 & C\,{\sc iv} & 1548.2041  & 0.12 & Tentative Identification\\ 
                                               &           & 5.7911 & C\,{\sc iv} & 1550.7812  & 0.06 & Tentative Identification\\
                                               &           & 2.8348 & Mg\,{\sc ii} & 2796.3543  & 0.18 & Tentative Identification\\
                                               &           & 2.8334 & Mg\,{\sc ii} & 2803.5315  & 0.22 & Tentative Identification\\
\\
SDSS\,J084035.09+562419.9    & 5.774  & 5.5940 & Si\,{\sc ii}  & 1526.70698  & 0.52  & \\
                                               &           & 5.5938 & Fe\,{\sc ii}  & 1608.45085  & 0.34  & \\
                                               &           & 5.5943 & Al\,{\sc ii}  & 1670.7886  & 0.44  & \\
\\
SDSS\,J103027.01+052455.0    & 6.28    & 5.7238 &  C\,{\sc iv} &  1548.2041 & 0.65 & \\
                                               &           & 5.7241 &  C\,{\sc iv} & 1550.7812  & 0.41 & \\
                                               &           & 5.8299: &  C\,{\sc iv} & 1548.2041  & \ldots & $z_{\rm abs}$ and $W_0$ uncertain\\
                                               &           & 5.8299: &  C\,{\sc iv} & 1550.7812  & \ldots & $z_{\rm abs}$ and $W_0$ uncertain\\
\\
SDSS\,J113717.73+354956.9    & 5.962  & 5.0127  & Al\,{\sc ii}  & 1670.7886  & 0.70  & \\
                                               &           & 5.0120 & Fe\,{\sc ii}  & 1608.45085  & 0.65  & \\
                                               &           & 2.6260 & Mg\,{\sc ii} & 2796.3543  & 0.52 & \\
                                               &           & 2.6261 & Mg\,{\sc ii} & 2803.5315  & 0.55 &\\
                                               &           & 5.7380 &  C\,{\sc iv} & 1548.2041  & 0.33 & \\
                                               &           & 5.7387 &  C\,{\sc iv} & 1550.7812  & 0.20 &\\
\\
SDSS\,J114816.64+525150.2    & 6.421  &  2.5077 & Mg\,{\sc ii} & 2796.3543  & 0.56 & Tentative Identification\\
                                               &           & 2.5084 & Mg\,{\sc ii} & 2803.5315  & 0.29 & Tentative Identification\\
                                               &           & 2.7727 & Mg\,{\sc ii} & 2796.3543  & 0.77 & \\
                                               &           & 2.7728 & Mg\,{\sc ii} & 2803.5315  & 0.60 & \\
                                               &           & 3.5545 & Fe\,{\sc ii} & 2344.2139  & 0.34 & \\
                                               &           & 3.5544 & Fe\,{\sc ii} & 2382.7652  & 0.96 & \\
\\
SDSS\,J130608.26+035626.3     & 5.99   & 2.5309 & Mg\,{\sc ii} & 2803.5315  & 2.50 & $\lambda 2796.3543$ outside spectral range$^{\rm c}$\\
\\
SDSS\,J160253.98+422824.9     & 6.051 & 2.7138 & Mg\,{\sc ii} & 2796.3543  & 0.44 & \\
                                               &           & 2.7135 & Mg\,{\sc ii} & 2803.5315  & 0.38 & \\
\\
SDSS\,J205406.49$-$000514.8   & 6.062  & 2.5973 & Mg\,{\sc ii} & 2796.3543  & 2.05 & \\
                                                &           & 2.5995 & Mg\,{\sc ii} & 2803.5315  & 2.29 & \\
\\
\hline
    \end{tabular}
    \smallskip

$^{\rm a}$From the compilation by Morton (2003).\\
$^{\rm b}$This value of equivalent width is in the observed frame (since the feature is unidentified).\\
$^{\rm c}$Both members of the doublet are present in the GNIRS spectrum published by Simcoe (2006).\\

     \label{tab:ew}
\end{minipage}
\end{table*}

\subsection{SDSS\,J083643.85+005453.3}
\label{sec:QSO3}

No absorption lines were identified in the 
$450$\,\AA\ stretch of the near-IR spectrum
of this QSO recorded with ISAAC (see bottom panel of
Figure~\ref{fig:spec3}).

\subsection{SDSS\,J084035.09+562419.9}
\label{sec:QSO4}

Our NIRSPEC spectrum of this QSO includes the 
broad \civ\ emission line at redshift 
$z_{\rm em} = 5.774$, to be compared with the 
SDSS value $z_{\rm em} = 5.85 \pm 0.02$ reported by Fan et al. (2006b).
We find no \civ\ systems,
but we do detect three absorption lines from a 
low ionisation system at $z_{\rm abs} = 5.5940$
(middle panel of Figure~\ref{fig:spec1}, and Table~\ref{tab:ew}).
It would be of great interest to determine
the metallicity (and other physical
parameters) of such high redshift absorbers
which may be associated with neutral gas in galaxies;
however, it is hard to envisage how this can be
accomplished, given that the associated Lyman series
lines are essentially indistinguishable against the 
backdrop of the nearly-black \lya\ forest.


 \begin{table*}
 \begin{minipage}{152mm}
 \caption{\textsc{Parameters of identified \civ\ absorption systems}}
   \begin{tabular}{llllll} 
\hline 
\hline
QSO & Instrument & $z_{\rm abs}$ & $\log N$(C\,{\sc iv})/cm$^{-2}$) & $b$ (\kms) & Comments\\ 
\hline 
SDSS\,J103027.01+052455.0 &  ISAAC      &   $5.7238 \pm 0.0001$  &  $14.39 \pm 0.05$ &  $61\pm 6$  &\\
SDSS\,J103027.01+052455.0 &  ISAAC      &   $5.8288 \pm 0.0002$  &  $14.45 \pm 0.13$ &  $26\pm 9$  &\\
SDSS\,J113717.73+354956.9 &  NIRSPEC  &   $5.7383 \pm 0.0002$  &  $14.20 \pm 0.15$ &  $30\pm 12$ & \\
\\
SDSS\,J081827.40+172251.8  &  ISAAC     &   $5.7899 \pm 0.0003$  &  $13.57 \pm 0.16$ &  $39\pm 25$  & Tentative identification\\
\\
\hline
     \label{tab:abs}
 \end{tabular}
 \end{minipage}
 \end{table*}

\subsection{SDSS\,J103027.01+052455.0}
\label{sec:QSO5}

This was the first QSO to be observed as part of this programme
and a portion of its spectrum was published by Ryan-Weber et al. (2006).
In Figure~\ref{fig:spec4} we show the full ISAAC spectrum 
which extends from 9880\,\AA\ to 10\,780\,\AA, recorded
with three, partially overlapping, grating settings.
We find two strong \civ\ doublets, at 
$z_{\rm abs} = 5.7238$ and 5.8288.
The latter was considered as marginal by Ryan-Weber et al. (2006),
but independent observations by Simcoe (2006) with the 
Gemini Near-Infrared Spectrograph (GNIRS) confirm
its reality. Even so, it is difficult with the quality of
our data to measure the equivalent widths of the 
two \civ\ lines in this absorber, as they appear to be 
blended with other broader features.

\subsection{SDSS\,J113717.73+354956.9 }
\label{sec:QSO6}

This is one of our best NIRSPEC spectra (see Figure~\ref{fig:spec1}).
The emission redshift we determine from C\,{\sc iv}\,$\lambda 1549.062$,
$z_{\rm em} = 5.962$, is lower than the 
value $z_{\rm em} = 6.01 \pm 0.02$ reported by Fan et al. (2006b) 
from the discovery SDSS spectrum.
We identify absorption lines from three separate systems
in the 1630\,\AA\ portion of the spectrum recorded with NIRSPEC:
a high redshift, low ionisation system at $z_{\rm abs} = 5.0124$,
a Mg\,{\sc ii} absorber at $z_{\rm abs} = 2.6261$,
and a \civ\ doublet at $z_{\rm abs} = 5.7383$.
The equivalent widths of all these lines are given in Table~\ref{tab:ew}.

\subsection{SDSS\,J114816.64+525150.2 }
\label{sec:QSO7}

This is the highest redshift QSO in the present sample.
We measure $z_{\rm em} = 6.421$ from the Si\,{\sc iv}\,$\lambda 1396.752$
broad emission line, in good agreement with the value
$z_{\rm em} = 6.42$ published by Fan et al. (2006c).
Our NIRSPEC spectrum (top panel of Figure~\ref{fig:spec2})
shows a definite Mg\,{\sc ii} doublet at $z_{\rm abs} = 2.7728$,
and a possible one at $z_{\rm abs} = 2.5080$ which
appears to be part of a broad feature 
of uncertain identity. We also cover two Fe\,{\sc ii}
lines, $\lambda\lambda 2344, 2382$, in a strong low ionisation 
system at $z_{\rm abs} = 3.5544$; 
other Fe\,{\sc ii} and Mg\,{\sc ii} lines in this system
can be recognised in the spectrum published by 
Barth et al. (2003).\footnote{We are grateful to the referee,
Xiaohui Fan, for pointing out the existence of this absorption
system.}

\subsection{SDSS\,J130608.26+035626.3}
\label{sec:QSO8}

The relatively short integration (two hours) of this object
with ISAAC resulted on a relatively noisy spectrum
(S/N\,$\simeq 5$) reproduced in the bottom panel of
Figure~\ref{fig:spec4} and already reported by
Ryan-Weber et al. (2006).
No \civ\ doublets are detected. The strong absorption line
present in the spectrum is the longer wavelength member of 
the Mg\,{\sc ii}\,$\lambda\lambda 2796.3543, 2803.5315$
doublet at $z_{\rm abs} = 2.5309$; both lines can be discerned
in the GNIRS spectra of this object obtained independently
by Simcoe (2006) and by Jiang et al. (2007).

\subsection{SDSS\,J160253.98+422824.9}
\label{sec:QSO9}

Our measured $z_{\rm em} = 6.051$ is in reasonable agreement
with $z_{\rm em} = 6.07$ from the SDSS (Fan et al. 2006c).
We detect a definite Mg\,{\sc ii} system at $z_{\rm abs} = 2.7137$,
as well as a puzzling broad absorption feature (see middle panel
of Figure~\ref{fig:spec2}). If this were 
a C\,{\sc iv} `mini-BAL', its ejection velocity relative to 
$z_{\rm em}$ would be 7150\,\kms.

\subsection{SDSS\,J205406.49$-$000514.8}
\label{sec:QSO10}

Even a relatively short exposure (5400\,s) with NIRSPEC
is sufficient to reveal the very strong Mg\,{\sc ii} doublet
present at $z_{\rm abs} = 2.5984$ (bottom panel of Figure~\ref{fig:spec2}).
This Mg\,{\sc ii} absorber, and the one at $z_{\rm abs} = 2.5309$
in SDSS\,J130608.26+035626.3, are close to the threshold for
classification as ``Ultra-Strong Systems'' in the SDSS survey
by Nestor, Turnshek, \& Rao (2005). It is interesting to note that
these authors find a steep redshift evolution in the frequency
of such systems, which become progressively rarer at $z \simlt 1$.

\section{Survey for intergalactic  C\,{\sc iv} at $z > 5$}

\subsection{Redshift Range Sampled}
In the last column of Table~\ref{tab:obs} we have listed
for each QSO the redshift intervals over which we searched
for \civ\ absorption. The values of $z_{\rm min}$,
which range from 5.16 to 5.52, correspond to wavelengths
close to the values of $\lambda_{\rm min}$ in column (5)
for the detection of $\lambda 1548.2041$. The values of
$z_{\rm max}$, which range from 5.66 to 6.20, 
correspond either to wavelengths close to 
the values of $\lambda_{\rm max}$ in column (5)
for the detection of $\lambda 1550.7812$, or to
 $z_{\rm em}- 3000$\,\kms, whichever is the lower
[the QSO emission redshifts are listed in column (3)].
We exclude `proximate systems' within 3000\,\kms\ of
the emission redshift because they may be associated with
the QSO environment, rather than with the more generally 
distributed IGM; however, none were found in the present
survey except for the two systems in  ULAS\,J020332.38+001229.2
which is not included in the
analysis anyway as it is a BAL QSO (see section~\ref{sec:QSO1}).

From these values of $z_{\rm min}$ and  $z_{\rm max}$
we calculate the corresponding absorption distances,
given by:
\begin{equation}
X(z)={2\over 3\Omega_{\rm M}}\,\{[\Omega_{\rm M}(1+z)^3+\Omega_\Lambda]^{1/2}-1\}
\end{equation}
\noindent which is valid for $\Omega_{\rm M} +\Omega_\Lambda = 1$,
where  $\Omega_{\rm M}$ and $\Omega_\Lambda$
are respectively the matter
and vacuum density parameters today.
With  $\Omega_{\rm M} = 0.3 $ and $\Omega_\Lambda = 0.7$,
we find that our survey covers a total absorption distance
$\Delta X = 25.1$.

\subsection{C\,{\sc iv} Column Densities}

Summarising the results of section~\ref{sec:QSOs},
we find three definite and one possible \civ\ doublets
over this absorption distance. 
We deduced values of column density
$\log N$(C\,{\sc iv})/cm$^{-2}$ for these absorbers
by fitting the observed absorption lines with
theoretical Voigt profiles generated by the 
{\sc vpfit} package.\footnote{{\sc vpfit} is available from
http://www.ast.cam.ac.uk/\textasciitilde rfc/vpfit.html}
For each absorber,  {\sc vpfit} determines  the most likely values of
redshift \za, Doppler width $b$ (km~s$^{-1}$),
and column density $\log N$(C\,{\sc iv})/cm$^{-2}$,
by minimizing the difference
between the observed line profiles and
theoretical ones 
convolved with the instrumental
point spread function.
Vacuum rest wavelengths and $f$-values of the
\civ\ transitions are from the compilation
by Morton (2003). 
For an illustrative use of this line-fitting software,
see Rix et al. (2007). 

Table~\ref{tab:abs} lists the values returned by
{\sc vpfit}, together with their $1 \sigma$ errors, 
for the four C\,{\sc iv} absorbers considered here.
Strictly speaking, these values of $N$(C\,{\sc iv})
are lower limits to the true column densities because
at the relative coarse spectral resolution of our data 
the absorption lines we see are likely to be the 
unresolved superpositions of several components
(e.g. Ellison et al. 2000), 
some of which may be saturated. 
This concerns applies particularly to the 
$z_{\rm abs} = 5.7383$ doublet in  SDSS\,J113717.73+354956.9 
for which the $b$ value on which {\sc vpfit} converged
($b = 30$\,\kms,
to be interpreted as an `equivalent' velocity
dispersion parameter describing the ensemble of
components) is much smaller 
than the NIRSPEC instrumental broadening
$b_{\rm instr} = 110$\,\kms---in other words,
these doublet lines are totally unresolved.
On the other hand, the doublet ratio 
$W(1548)/W(1550) = 1.65$ is consistent with
only a moderate degree of line saturation.
In the other three cases, the equivalent $b$-values
are comparable to, or larger than, $b_{\rm instr} = 32$\,\kms\ 
of ISAAC (see Table~\ref{tab:abs}), indicating
that the \civ\ absorption lines are partially
resolved. For the remainder of our analysis,
we shall assume that the column densities of
\civ\ have not been underestimated by large
factors, although this assumption will need to be
verified in future with higher resolution observations.

\subsection{Mass density of \civ}

It has become customary to express the mass density
of an ion (in this case \civ) as a fraction of the critical 
density today:
\begin{equation}
\Omega_{\rm C\,\textsc{iv}} = 
\frac{H_0 \, m_{\rm C\,\textsc{iv}}}{c \, \rho_{\rm crit}}\int N f(N) {\rm d}N, 
\label{eq:O_CIVa}
\end{equation}
where $H_0 = 100\,h$\,\kms~Mpc$^{-1}$ is the Hubble constant,
$m_{\rm C\,\textsc{iv}}$ is the mass of a \civ\ ion, $c$ is the speed of light,
$\rho_{\rm crit} = 1.89 
\times 10^{-29}\,h^2$\,g~cm$^{-3}$,
and $f(N)$ is the number of \civ\
absorbers per unit column density per unit 
absorption distance $X(z)$. 
In the absence of sufficient statistics to recover the column density
distribution function $f(N)$,
as is the case here, the integral in eq.~(\ref{eq:O_CIVa})
can be approximated by a sum:
\begin{equation}
\int N f(N) {\rm d}N = 
\frac{\sum_{i} N_{i} {\rm (C\,\textsc{iv})}}{\Delta X}
\label{eq:O_CIVb}
\end{equation}
with an associated fractional variance:
\begin{equation}
\left (
\frac{\delta\Omega_{\rm C\,\textsc{iv}}}{\Omega_{\rm C\,\textsc{iv}}} \right )^2 = 
\frac{\sum_{i} [N_{i}{\rm (C\,\textsc{iv})}]^2}
{\left [ \sum_{i} N_{i} {\rm (C\,\textsc{iv})} \right ] ^2},
\label{eq:O_CIVc}
\end{equation}
as proposed by Storrie-Lombardi, McMahon, \& Irwin (1996).
With $h = 0.70$, eqs.~(\ref{eq:O_CIVa}) and (\ref{eq:O_CIVb}) then lead to:
\begin{equation}
\Omega_{\rm C\,\textsc{iv}} = 1.63 \times 10^{-22}~ 
\frac{\sum_{i} N_{i}{\rm (C\,\textsc{iv})}}{\Delta X}.
\label{eq:O_CIVd}
\end{equation}

\begin{figure}
\vspace{-0.35cm}
\centerline{\hspace{0.1cm}
\includegraphics[width=0.99\columnwidth,clip,angle=0, viewport=75 300 510 610]{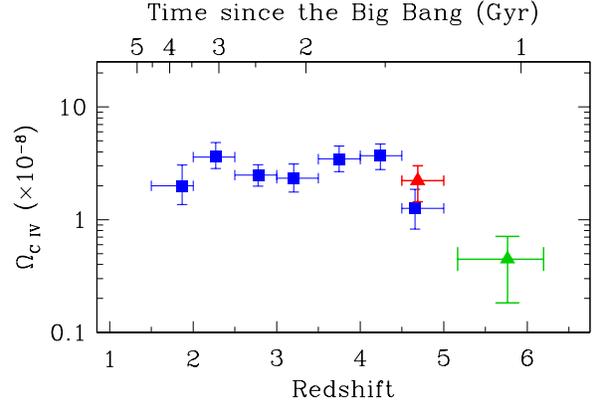}}
\vspace{0.15cm}
\caption{Cosmological mass density of \civ\ as a function of redshift.
  The blue squares show the measurements by Songaila (2001), the
  red triangle is from Pettini et al. (2003),
  and the green triangle is the value deduced here. 
  All values plotted have been reduced to the `737'  cosmology adopted
  in the present work. Error bars are $1 \sigma$.
  While this plot shows the actual values of $\Omega_{\rm C\,{\textsc iv}}$
  measured, they are not strictly comparable because each 
  of the three surveys had a different sensitivity limit.
 This issue is discussed in detail in the text (section~\ref{sec:complete}).
 }
\label{fig:O_CIVa}
\end{figure}

\noindent Summing up the values of $N_{i}$(C\,{\sc iv}) in Table~\ref{tab:abs}
and dividing by $\Delta X = 25.1$, we find:
\begin{equation}
\Omega_{\rm C\,\textsc{iv}} = (4.4 \pm 2.6) \times 10^{-9}
\label{eq:O_CIVe}
\end{equation}
over the redshift interval $z_{\rm abs} = 5.2$--6.2
at a mean $\langle  z_{\rm abs} \rangle = 5.76$.
We have not included the possible system at $z_{\rm abs}=5.7899$
in the summation; its inclusion would increase  
$\Omega_{\rm C\,\textsc{iv}}$ by only $5\%$.
For comparison, Pettini et al. (2003) measured
$\Omega_{\rm C\,\textsc{iv}} = (2.2 \pm 0.8) \times 10^{-8}$
at $\langle  z_{\rm abs} \rangle = 4.69$ 
(in the same cosmology as adopted here); 
these two values, together with the lower redshift 
determinations by Songaila (2001), are plotted
in Figure~\ref{fig:O_CIVa}.

Thus, at first sight, $\Omega_{\rm C\,\textsc{iv}}$
seems to have dropped by about a factor of five
in only $\sim 300$\,Myr, as we move back in time
from $z \sim  4.7$ to $z \sim 5.8$. 
However, before
we can be confident of such a rapid build-up of 
\civ\ at these early epochs in the Universe
history (these redshifts correspond to only
$\sim 1$\,Gyr after the Big Bang---see top axis
of Figure~\ref{fig:O_CIVa}), we have to look
carefully at the different completeness limits
of the surveys whose results are collected in 
Figure~\ref{fig:O_CIVa}. 
In particular, the sensitivities of absorption line
searches at near-IR wavelengths, 
such as the one reported here, are still 
lower than those achieved at optical wavelengths
with comparable observational efforts and we suspect that
this difference must be contributing to some extent
to the high redshift drop evident in Figure~\ref{fig:O_CIVa}.


\section{Completeness Limit of the Survey}
\label{sec:complete}

In order to quantify this effect, we conducted a series of
Monte Carlo simulations, as follows.
We used {\sc vpfit} to generate theoretical \civ\ doublets for
ranges of values of $b$ and $N$(C\,\textsc{iv}); these fake
absorption lines were then inserted at random redshifts
into the real spectra, after 
convolution with the appropriate instrumental profile
(depending on whether the spectrum to be `contaminated'
was a NIRSPEC or an ISAAC one) and with random noise reflecting
the measured S/N of the spectrum at that wavelength.
We then tested whether we could recover these \civ\
doublets by visual inspection of the spectra, in the same
way as the real absorbers were identified.
So as to test the sensitivity of our survey as comprehensively
as practical, we chose three sets of values for the 
velocity dispersion parameter, $b = 20$, 40 and 60\,\kms,
reflecting the range of $b$-values of the real
\civ\ doublets (see Table~\ref{tab:abs}).
For the column density we considered twelve values,
in 0.1\,dex steps between $\log N({\rm C\,\textsc{iv})/cm}^{-2} = 13.4$ and 14.5.
For each combination of $b$ and $\log N({\rm C\,\textsc{iv}})$,
we created 40 Monte Carlo realisations, so that the full
series of completeness tests involved 1440 fake \civ\ doublets 
distributed within our parameter space.


\begin{figure}
\vspace{-0.35cm}
\centerline{\hspace{0.1cm}
\includegraphics[width=0.99\columnwidth,clip,angle=0, viewport=75 300 510 610]{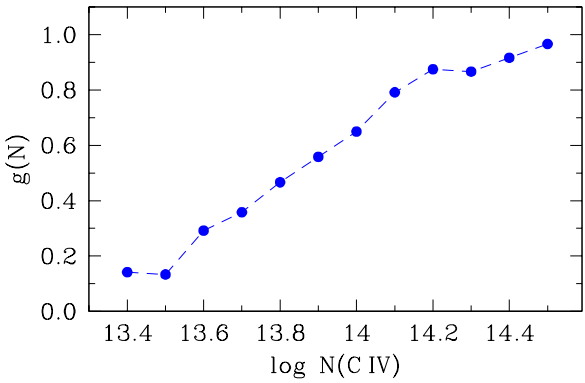}}
\vspace{0.15cm}
\caption{Sensitivity function to \civ\ doublets in our survey,
deduced from 1440 Monte Carlo realisations. See text (section~\ref{sec:complete})
for further details.
 }
\label{fig:gN}
\end{figure}

We found that the fraction of \civ\ doublets recovered, $g(N)$, did not
depend sensitively on $b$ nor $z_{\rm abs}$ but, as expected, was
a function of column density---see Figure~\ref{fig:gN}. 
While we seem to be $\sim 90\%$ complete for column densities
$\log N({\rm C\,\textsc{iv})/cm}^{-2} \geq 14.2$, our sensitivity
drops fairly rapidly below this value: we can only detect about half
of the \civ\ doublets with half this column density and, if 
$N$(C\,\textsc{iv}) drops by a further factor of two, 
we miss some 85\% of the absorbers. 
For comparison, the estimate of
$\Omega_{\rm C\,\textsc{iv}}$ by Pettini et al. (2003)
refers to \civ\ absorbers with 
$\log N({\rm C\,\textsc{iv})/cm}^{-2} \geq 13.0$ 
(and includes a 22\% correction for the fall in completeness of 
their Keck-ESI spectra from
$\sim 100\%$ at $\log N({\rm C\,\textsc{iv})/cm}^{-2} \geq 13.3$
to 45\% at $\log N({\rm C\,\textsc{iv})/cm}^{-2} = 13.0$).
Songaila's (2001) estimates of $\Omega_{\rm C\,\textsc{iv}}$,
on the other hand, refer to the yet wider column density interval
$12.0 \leq  \log N({\rm C\,\textsc{iv})/cm}^{-2} \leq 15.0$
which she could access thanks to the high spectral resolution
and S/N of her Keck-HIRES spectra. From her Figure~1,
it appears that her number counts become progressively
less complete below $\log N({\rm C\,\textsc{iv})/cm}^{-2} \simeq 13.0$,
although no correction was applied to account for this.

From the above discussion it is evident that, in order to assess
the significance of the suggested drop in 
$\Omega_{\rm C\,\textsc{iv}}$ beyond $z \simeq 4.7$,
we need to consider the results of these different surveys over the 
same range of \civ\ column densities.
Referring to Figure~\ref{fig:O_CIVa}, we see that
we are $>50\%$ complete for 
$\log N({\rm C\,\textsc{iv})/cm}^{-2} \geq 13.8$, so that
we can compare with the surveys by Songaila (2001)
and Pettini et al. (2003) in this column density regime.
Specifically, we have adjusted the values
$\Omega_{\rm C\,\textsc{iv}}$ reported by those authors
as follows:
\begin{equation}
\Omega^{\prime}_{\rm C\,\textsc{iv}} =
\Omega_{\rm C\,\textsc{iv}} \times
\frac{\int^{{10}^{15.0}}_ {{10}^{13.8}}N f(N) {\rm d}N}{\int^{{10}^{15.0}}_ {{10}^{12.0}}N f(N) {\rm d}N}
\label{eq:O_CIVcorr1}
\end{equation}
for the values published by Songaila (2001), and
\begin{equation}
\Omega^{\prime}_{\rm C\,\textsc{iv}} =
\Omega_{\rm C\,\textsc{iv}} \times
\frac{\int^{{10}^{15.0}}_ {{10}^{13.8}}N f(N) {\rm d}N}{\int^{{10}^{15.0}}_ {{10}^{13.0}}N f(N) {\rm d}N}
\label{eq:O_CIVcorr2}
\end{equation}
for the value published by Pettini et al. (2003),  assuming
\begin{equation}
f(N) = 10^{-12.65} N_{13}^{-1.8} ,
\label{eq:fN}
\end{equation}
where $N_{13}$ is $N$(C\,{\sc iv}) in units of
$10^{13}$\,cm$^{-2}$, as determined by Songaila (2001)\footnote{Again,
adjusted to the present cosmology.}.

\begin{figure}
\vspace{-0.35cm}
\centerline{\hspace{0.1cm}
\includegraphics[width=0.99\columnwidth,clip,angle=0, viewport=75 300 510 610]{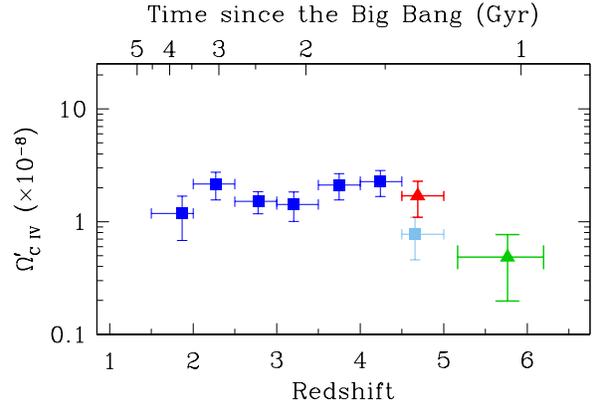}}
\vspace{0.15cm}
\caption{Cosmological mass density of \civ\ as a function of redshift,
  taking into account only \civ\ absorbers with
  column densities
  $13.8 \leq \log N({\rm C\,\textsc{iv})/cm}^{-2} \leq 15.0$,
  to which all three surveys considered here are sensitive.
  The symbols have the same meaning as in Figure~\ref{fig:O_CIVa}.
  For the highest redshift data point in the sample by Songaila (2001),
  shown by the pale blue square, the correction of eq.~(\ref{eq:O_CIVcorr1}) 
  probably underestimates the value of $\Omega^{\prime}_{\rm C\,\textsc{iv}}$.
 }
\label{fig:O_CIVb}
\end{figure}

The lower limits of integration in the denominators
of eqs.~(\ref{eq:O_CIVcorr1}) and (\ref{eq:O_CIVcorr2})
reflect the sensitivity limits of the corresponding surveys. 
Since the survey by Songaila (2001) is only partially
complete for \civ\ absorbers with 
$12.0 \leq \log N({\rm C\,\textsc{iv})/cm}^{-2} \leq 13.0$,
eq.~(\ref{eq:O_CIVcorr1}) may overestimate the
correction required---this works in the sense of
diluting the evidence for a drop in
$\Omega_{\rm C\,\textsc{iv}}$ at higher redshifts.
The effect is most pronounced for the highest
redshift bin in Songaila's sample, because 
her data is least sensitive to column densities 
$\log N({\rm C\,\textsc{iv})/cm}^{-2} < 13.0$
at $z_{\rm abs} \geq 4.5$
(see Figure~2 of Songaila 2001); accordingly 
we have indicated her highest redshift data point with a different
colour in Figure~\ref{fig:O_CIVb}.

For the NIRSPEC+ISAAC sample presented here,
we have modified eq.~(\ref{eq:O_CIVd}) to include our
sensitivity function $g(N)$ shown in Figure~\ref{fig:gN}:
\begin{equation}
\Omega^{\prime}_{\rm C\,\textsc{iv}} = 1.63 \times 10^{-22}~ 
\frac{\sum_{i} N_{i}{\rm (C\,\textsc{iv})} \times g(N_i)}
{\Delta X}.
\label{eq:O_CIVcorr3}
\end{equation}
All the errors have been scaled accordingly, so as to maintain
the same percentage error.


Values of $\Omega^{\prime}_{\rm C\,\textsc{iv}}$
are plotted in Figure~\ref{fig:O_CIVb}.
Evidently, we still see a drop in the mass density of \civ\ as we look
back beyond $z \simeq 4.7$, albeit at a reduced significance
compared to Figure~\ref{fig:O_CIVa}. 
Considering only absorbers with 
$\log N({\rm C\,\textsc{iv})/cm}^{-2} \geq 13.8$,
we find that $\Omega^{\prime}_{\rm C\,\textsc{iv}}$ 
decreases by factor of $\sim 3.5$, as we move from
$\Omega^{\prime}_{\rm C\,\textsc{iv}} = (1.7 \pm 0.6) \times 10^{-8}$
at $\langle z \rangle = 4.69$ to 
$\Omega^{\prime}_{\rm C\,\textsc{iv}} = (5 \pm 3) \times 10^{-9}$
at $\langle z \rangle = 5.76$ ($1 \sigma$ errors).
The drop is significant at the $4 \sigma$ level.

\section{Absorption Line Statistics}

We can examine the evidence for a decrease in intergalactic \civ\
at redshifts $z > 4.7$ in an alternative way from considering simply
the sum of their column densities. It is instructive to ask how
the number of detected \civ\ doublets 
compares with the number we may have expected under the
assumption of an invariant column density distribution, $f(N)$.
Again, we assume $f(N) = 10^{-12.65} N_{13}^{-1.8}$ (Songaila 2001),
which implies $f(>N) =2.8\,N_{13}^{-0.8}$ per unit absorption distance.

Our survey is not uniform in either S/N nor resolution, and both
factors determine the minimum value of $N$(C\,{\sc iv}) detectable.
We therefore calculated the expected value of $f(>N)$ on a 
spectrum by spectrum basis, using as threshold the $5 \sigma$ detection
limit for the equivalent width of an absorption line:
\begin{equation}
W_0(5 \sigma) = \frac{5 \delta \lambda}{1+z} \times \frac{1}{\rm{S/N}}
\label{eq:W5s}
\end{equation}
where $\delta \lambda$ is the resolution in \AA.
Values of resolution and S/N for each spectrum are listed in Table~\ref{tab:obs}.
With $W_0(5 \sigma)$ is associated a minimum column density:
\begin{equation}
N_{\rm C\,\textsc{iv}}(5 \sigma) = 1.13 \times 10^{20}  \times 
\frac{W_0(5 \sigma)}{\lambda^2 f}~ {\rm cm}^{-2}
\label{eq:NCIV}
\end{equation}
where $\lambda$ and $W_0$ are both in \AA\ and $f$ is the oscillator
strength of the transition (in this case C\,{\sc iv}$\lambda 1550.7812$,
the weaker member of the doublet). 

\begin{figure}
\vspace{-0.25cm}
\centerline{\hspace{0.1cm}
\includegraphics[width=0.99\columnwidth,clip,angle=0]{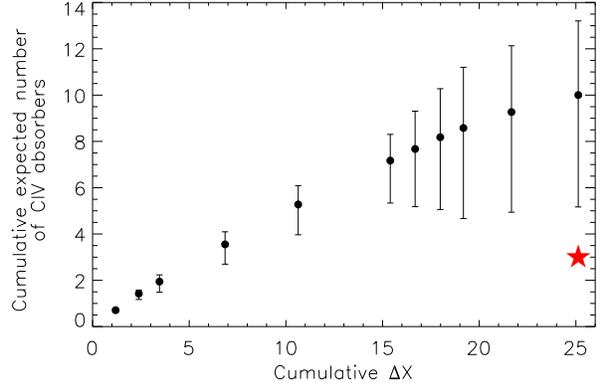}}
\caption{Cumulative number of \civ\ absorption systems expected from
  our survey (if there is no evolution in the column density
  distribution function from redshift $z \sim 3$ to 6). The cumulative
  number is plotted as a function of the survey pathlength, $\Delta X$,
  which increases as each spectrum is added to data set. We expect to
  find 10 \civ\ absorption systems in our total pathlength of
  $\Delta X = 25.1$. The star symbol denotes the three 
  absorbers we detect in our entire survey in the redshift
  range $5.2<z<6.2$. }
\label{fig:cum_ENS}
\end{figure}

In Figure~\ref{fig:cum_ENS} we plot the cumulative number of absorbers
expected (for the minimum column density 
to which each spectrum is sensitive) as a
function of the cumulative pathlength. 
Over our total survey
pathlength, $\Delta X = 25.1$,
we expect to detect $10^{+3}_{-5}$ \civ\ absorbers.
We instead find three absorbers,
indicating that the distribution of \civ\ column densities
which applies at redshift $1.5 \leq z \leq 4.5$ does \emph{not}
continue unchanged to $ 5.2 \leq z \leq 6.2$.
The likelihood of detecting three \civ\ systems
when ten are expected is less than 1\%.
It is interesting to note that the absorption
systems we do detect were found in spectra with column density
sensitivities in the middle of the range of our survey:
$ N_{\rm C\,\textsc{iv}}(5 \sigma) = 9.1 \times 10^{13}$\,cm$^{-2}$ 
and $1.2 \times 10^{14}$\,cm$^{-2}$ for
SDSS\,J113717.73+354956.9 and SDSS\,J103027.01+052455.0
respectively. The absorber towards SDSS\,J113717.73+354956.9
represents a $6.6 \sigma$ detection. Thus, it seem unlikely that
we are missing a considerable number of absorption systems above our
$5 \sigma$ sensitivity limit.


\section{Discussion}

To summarise, by extending our previous searches for intergalactic
\civ\ to near-IR wavelengths, we have found that
both the number of such absorbers per unit absorption
distance and the comoving  mass of triply ionised carbon
are approximately three times lower at $z  \simeq 5.8$
than at $z < 4.7$. Although our results refer to 
the stronger \civ\ systems in the column
density distribution function, 
with $\log N{\rm (C\,\textsc{iv})/cm}^{-2} \geq 13.8$,
the evolution appears to extend at least to absorbers
with half this column density (Becker et al. 2009).

Before considering the implications of these results, we 
reflect briefly on possible caveats. As already mentioned,
we may have underestimated the column densities
of \civ\ through line saturation; we do not expect this 
effect to be so severe as to cause (spuriously) the fall
in \Ociv\ evident in Figure~\ref{fig:O_CIVb},  but only
higher resolution observations will quantify the corrections
required (if any). It must also be borne in mind
that the distribution of \civ\ at $z > 5.2$ seems to be
extremely patchy: of the nine sight-lines sampled, seven
show no absorption, while two absorbers are found in
the direction of one QSO. Thus, our statistics are very
prone to what in galaxy counts is referred to as 
`cosmic variance'. 
This can explain why the pilot observations
by Ryan-Weber et al. (2006) suggested a higher
value of \Ociv\---the first stage of
our survey fortuitously found one-third of the
$\sum_{i} N_{i} \rm (C\,\textsc{iv})$ of the 
whole sample presented here, even though it covered only 
$\sim 15\%$ of the total $\Delta X$.
Conversely, Becker et al. (2009) detected \emph{no} absorbers
in their higher resolution NIRSPEC spectra of four QSOs (three
in common with our sample) over less than half of the 
$\Delta X$ of our survey.
In future, better statistics will undoubtedly improve the 
accuracy of determinations of  \Ociv\ at $z > 5$;
for the moment we shall proceed on the assumption that
our estimate of the error on this quantity, as shown in
Figure~\ref{fig:O_CIVb}, is realistic.

The downturn in \Ociv\ we have discovered as we approach
$z \sim 6$ is clear evidence of important changes taking
place in the Universe at, or before, this epoch. 
In the following discussion we shall find it more convenient
to `turn the time arrow around', and to think of the evolution
as time progresses from higher to lower redshifts. 
The rise in \Ociv\ we see in Figure~\ref{fig:O_CIVb} 
could result either from an increase in the amount
of intergalactic carbon, or from changes in the
ionisation balance favouring the fraction of carbon
which is triply ionised, or both. Indeed, there are indications
from previous work that both effects may be at play.

By integrating the galaxy luminosity function down to
1/5 of the fiducial luminosity at $z = 3$ (that is,
considering galaxies with $L \geq 0.2\,L^{\ast}_{z = 3}$),
Bouwens et al. (2008) found that the cosmic star-formation
rate density increased by a factor of $\simgt 4$ 
over the 400\,Myr period from
$z = 9$ to $z = 6$. The growth in intergalactic \civ\
we see at $z < 6$ may well be a reflection of this
enhanced star-formation activity, as the products of 
stellar nucleosynthesis accumulate in the IGM, presumably
transported by galactic outflows. Note that this is carbon
produced by massive stars, with evolutionary time-scales
of less than 10\,Myr (Pettini et al. 2008b; Pettini 2008);
it is therefore quite conceivable that over $\sim 400$\,Myr
the products of Type\,II supernovae could have travelled
considerable distances from their production sites.

On the other hand, Fan et al. (2006c) have argued for a 
marked increase in the transmission of the \lya\ forest 
at $z \simlt 6$, which has been interpreted as marking 
the overlap of the hydrogen ionisation fronts from
star-forming galaxies, as the IGM became transparent 
to Lyman continuum photons (e.g. Gnedin \& Fan 2006).
Opinions are still divided, however, as to the correct
interpretation of the \lya\ forest evolution at these
high redshifts (e.g. Becker, Rauch, \& Sargent 2007 and
references therein). Coupled with the rise in the
comoving density of bright QSOs (Fan 2006),
a higher IGM transmission would quickly ionise 
a larger fraction of carbon into \civ.

To make progress beyond these qualitative remarks,
detailed modelling is necessary. As mentioned in the
Introduction, B. Oppenheimer and collaborators have
devoted considerable attention to this problem and shown,
in particular, how momentum-driven winds from star-forming
galaxies can reproduce the behaviour of \Ociv\ at 
redshifts $z < 5$. It is therefore of interest to consider their
model predictions for earlier epochs.
Their latest models (Oppenheimer, Dav{\'e}, \& Finlator 2009)
do indeed entertain a marked
growth in \Ociv\ from $z = 8$ to 5, driven by \emph{both} effects
mentioned above. In their scenario, the overall
density of carbon increases by a factor of $\sim 2$
between $z = 5.8$ and 4.7.
The accompanying rise in  \Ociv\ depends on the 
spectral shape of the radiation field which dominates
the ionisation equilibrium. Oppenheimer et al. (2009)
consider two possibilities: the metagalactic background
in the formulation by Haardt \& Madau (1996) with
recent updates, and 
a more local field dominated by early-type stars
in the nearest galaxy to the \civ\ absorbers, in recognition
of the fact that cosmic reionisation was likely to be 
very patchy at these early epochs. The former produces
the larger change in \Ociv\ (for the reasons explained above),
which grows by a factor of $\sim 5$ between 
$z = 5.8$ and 4.7; the latter results in a more modest
rise,  by a factor of $\sim 2-3$.
 
Given our warnings about the effects of `cosmic variance' on
the current observational determinations of \Ociv,
it would be quite premature to compare in detail these
theoretical predictions with the data in Figure~\ref{fig:O_CIVb}
and, for example, draw conclusions on the shape of the 
ionising background at these early epochs. It is encouraging,
nevertheless, that models and data concur in a general trend
towards a build-up of \civ\ over the relatively short
cosmic period between $z \sim 6$ and 5. 
Two ingredients of the simulations by Oppenheimer 
et al. (2009) need to be clarified, however.
The first, and more important one, is that in their scenario 
the galaxies responsible for most of the IGM enrichment are 
of low mass and with luminosities well
\emph{below} the current observational limit 
$L \geq 0.2\,L^{\ast}_{z = 3}$. 
We know very little, if anything at all, about the galaxies
at the faint end of the luminosity function at $z > 5$, and this situation
is unlikely to be remedied until the launch of the
James Webb  Space Telescope in the next decade.
A second concern is the resonant opacity of the IGM at the frequencies
of the He\,{\sc ii} Lyman series which has been neglected until
recently, and which may
affect the distribution of carbon among its highly ionised states
(Madau \& Haardt 2009).

\subsection{Reionization}

The average metallicity of the IGM at a given epoch
reflects the accumulation of the heavy elements
synthesised by previous generations of stars
and expelled from their host galaxies.
The massive stars which explode as Type\,II supernovae
and seed the IGM with metals are also the sources of  
nearly all of the Lyman continuum photons produced by a burst of
star formation.  Consequently, it is relatively straightforward
to link a given IGM metallicity with a minimum number of
LyC photons which must have been produced up until that time.
The close correspondence between
the sources metals and photons makes the conversion
from one to the other largely independent 
of the details of the stellar initial mass function (IMF).
For example, Madau \& Shull (1996) calculate
that changing the index of a power-law formulation
of the IMF, such as Salpeter's (1955), from $\alpha = 2$
to 3, results in only a 10\% difference in the 
conversion between metallicity 
and number of LyC photons emitted.

When in the past this line of reasoning has been applied
to measures of the IGM metallicity at $z \sim 3$ 
(e.g. Madau \& Shull 1996; Miralda-Escude \& Rees 1997),
it has been concluded that the star formation activity
prior to that epoch probably produced sufficient photons to
reionise the Universe. With our new estimate of 
\Ociv, much closer in time to what may have been
the end of the process of reionisation, we are able to
revisit this question more critically.

We begin by calculating the IGM metallicity
by mass, $Z_{\rm IGM}$, implied by
our measure of \Ociv:
\begin{equation}
Z_{\rm IGM} = \frac{\Ociv}{\Omega_{\rm b}} 
\cdot \frac{\rm C_{\rm {TOT}}}{\rm C\,{\textsc{iv}}}
\cdot \frac{1}{A_{\rm C}}
\label{eq:Z_IGM1}
\end{equation}
where $\Omega_{\rm b} = 0.0224/h^2$ (Pettini et al. 2008a)
is the contribution of baryons to the critical density,
$\rm C\,{\textsc{iv}}/\rm C_{\rm {TOT}}$ is the
fraction of carbon which is triply ionised, and
$A_{\rm C}$ is the mass fraction of metals in carbon.
For comparison, in the Sun, $Z_{\odot} = 0.0122$
and $A_{\rm C} = 0.178$ (Asplund, Grevesse, \& Sauval 2005);
here we assume that the same value of $A_{\rm C}$
applies at high redshifts because: (a) the ratio of carbon to oxygen
(the major contributor to $Z_{\odot}$) is approximately
solar at low metallicities (Pettini et al. 2008b; Fabbian et al. 2009);
and (b) any departures are in any case likely to result in only 
a small correction compared to other uncertainties in the 
following discussion. 
We also adopt the conservative limit 
${\rm C\,\textsc{iv}}/{\rm C_{\rm {TOT}}} \leq 0.5$,
as done by Songaila (2001); this is the maximum
fractional abundance reached by C\,{\sc iv} under
the most favourable ionisation balance conditions.
With these parameters, our measured 
\begin{equation}
\Omega_{\rm C\,\textsc{iv}} = (4.4 \pm 2.6) \times 10^{-9}
\end{equation}
implies 
\begin{equation}
Z_{\rm IGM} = (1.1 \pm 0.6) \times 10^{-6} = (9 \pm 5) \times 10^{-5}\,Z_{\odot} .
\label{eq:Z_IGM2}
\end{equation}

\noindent In recent years, much theoretical attention has been directed
to estimating the likely properties of metal-poor early-type stars.
Schaerer (2002) pointed out that stellar luminosities at far-UV 
wavelengths are higher at lower metallicities, leading to a greater
output of LyC photons associated with stellar nucleosynthesis.
Specifically, the energy emitted in hydrogen ionising photons 
(per baryon) is related to the average cosmic metallicity by:
\begin{equation}
E_{\rm Z} = \eta \, m_{\rm p}c^2 \, \langle Z \rangle \, {\rm MeV}
\label{eq:E_Z}
\end{equation}
where $m_{\rm p}c^2 = 938$\,MeV is the rest mass of the proton
and $\eta$ is the stellar efficiency conversion factor.
Schaerer (2002) estimated $\eta = 0.014$ for stars with
$Z = 1/50 Z_{\odot}$, a factor of $\sim 4$  higher than at
solar metallicities (see also Venkatesan \& Truran 2003).
For an average energy of 21\,eV per LyC photon (Schaerer 2002),
our determination of 
$\langle  Z_{\rm IGM} \rangle = 1.1 \times 10^{-6}$
then implies that less than one ($\sim 0.7$) LyC photon per baryon
was emitted by early-type stars prior to $z = 5.8$. 

This background is clearly insufficient to keep the Universe
ionised.
When we take into account that: 
(a) an unknown fraction, $f_{\rm esc} < 1$,  of the emitted LyC photons 
will escape from the regions of star formation 
into intergalactic space, and (b) more than one 
intergalactic LyC photon per baryon is likely to be required 
to keep the clumpy IGM ionised against radiative recombinations
(e.g. Madau, Haardt, \& Rees 1999; Bolton \& Haehnelt 2007),
we reach the conclusion that our shortfall is by at least 
one order of magnitude. 

There are a number of possible solutions to this problem.
The most obvious one is to recall that
eq.~(\ref{eq:Z_IGM2}) gives the \emph{lower limit} to
the IGM metallicity, for two reasons. 
First,  our accounting 
of \Ociv\ only includes high column density absorbers,
with $N{\rm (C\,{\textsc iv})} \simgt 10^{14}$\,cm$^{-2}$.
A steepening of the column density distribution, $f(N)$,
at $z > 5.2$ could hide significant amounts of metals in lower
column density systems. However, the recent null result
by Becker et al. (2009) runs counter to this explanation.

Second, if ${\rm C\,\textsc{iv}}/{\rm C_{\rm {TOT}}} \ll 0.5$,
some of the tension would be relieved.
For example, in the models considered
by Oppenheimer et al. (2009) most of the C is doubly-ionised
at $z = 5$--6, and 
${\rm C\,\textsc{iv}}/{\rm C_{\rm {TOT}}} \approx 0.1$,
a factor of five lower than the upper limit assumed
in eq.~(\ref{eq:Z_IGM2}). On the other hand,
there are limits to the extent to which the shortfall
can be explained by ionisation effects alone. 
If the IGM is not highly ionised at these redshifts,
so that ${\rm C\,\textsc{iv}}/{\rm C_{\rm {TOT}}} \ll 0.5$,
we would also expect a non-negligible fraction of C to be singly
ionised. Unlike C\,{\sc iii}, which is virtually impossible
to observe at these redshifts because its resonance lines
are either too weak or  buried in the Ly$\alpha$ forest,  
C\,{\sc ii} has a well-placed strong transition
at rest wavelength $\lambda_0 =1334.5323$\,\AA\
which should provide, in conjunction with $N$(C\,\textsc{iv}),
some indication of the ionisation balance of C. 
For two of the C\,{\sc iv} absorbers in the present sample,
at $z_{\rm abs} = 5.7238$ and 5.8288 in line to 
SDSSJ103027.01+052455.0 (Table~\ref{tab:abs}),
the corresponding C\,{\sc ii}\,$\lambda 1334.5323$ lines
are covered by the ESI spectrum of Pettini et al. (2003).
In neither case is C\,{\sc ii} absorption detected;
the corresponding upper limits are
$N$(C\,{\sc ii})/$N$(C\,{\sc iv})\,$\leq 0.2$ at
$z_{\rm abs} = 5.7238$ and 
$N$(C\,{\sc ii})/$N$(C\,{\sc iv})\,$\leq 0.05$ at
$z_{\rm abs} = 5.8288$.
Clearly these are high-ionisation absorbers
for which it seems unlikely that C\,{\sc iv}
is only a trace ion stage of carbon.
In future, it will be possible to quantify
better the ionisation state of the metal-bearing
IGM at these redshifts with observations
targeted at other elements, such as Si which 
may be observable in three ion stages: Si\,{\sc ii},
Si\,{\sc iii}, and Si\,{\sc iv}.

Returning to $Z_{\rm IGM}$, another reason why the
value in eq.~(\ref{eq:Z_IGM2}) may
underestimate the cosmic mean metallicity is that it
does not include carbon in stars. The ensuing upward correction
however, is unlikely to be more than a factor of $\sim 2$,
unless current estimates (e.g. Bouwens et al. 2008) 
still miss a significant
fraction  of the cosmic star formation activity
prior to $z = 5.8$ as argued, for example, by
Faucher-Gigu{\`e}re et al. (2008).
It is also intriguing to note that Schaerer (2002) proposed
an efficiency factor as high as $\eta = 0.065$ 
(nearly five times higher than the value used here
in eq.~\ref{eq:E_Z}) for metal-free stars.
Additionally, the number of LyC photons 
in the IGM may be boosted
by a population of faint QSOs yet to be detected.

Possibly, all of these factors contribute and,  
given all the unknowns, 
one may argue that this kind of accounting 
exercise is premature.
Future work, aimed at improving the statistics
of \civ\ absorption at high redshifts and at
identifying the galaxies from which these metals may have originated,
will undoubtedly place the above speculations on much 
firmer observational ground.

\section{Conclusions}

We have conducted the largest survey to date for 
intergalactic metals
at redshifts $5.2 < z < 6.2$. 
Our sample consists of near-IR spectra of
nine high-redshift QSOs covering  
a total absorption distance $\Delta X=25.1$, seven times
greater than the pilot observations by Ryan-Weber et al. (2006).
Our main results can be summarised as follows:

1. We detect three definite and one possible \civ\ doublets;
the three definite cases all correspond to column densities
of triply ionised carbon in excess of $10^{14}$\,cm$^{-2}$.
We detect two further \civ\ doublets in a tenth QSO
which is not included in the final sample as it shows evidence
for mass ejection (a BAL QSO); in any case
these two systems are probably associated with the QSO
environment rather than being truly intergalactic.

2. From the three definite detections, we deduce a comoving
mass density 
$\Omega_{\rm C\,\textsc{iv}} = (4.4 \pm 2.6) \times 10^{-9}$
at a mean $\langle z \rangle = 5.76$.
This value represents a drop in 
$\Omega_{\rm C\,\textsc{iv}}$ by a factor of 
$\sim 3.5$ compared to $z \leq 4.7$ at the 
$4 \sigma$ significance level, after the different
levels of completeness of optical and near-IR surveys 
are taken into account.

3. If the column density distribution of \civ\ absorbers
remained invariant at $z > 4.7$ in its shape and normalisation
as determined at $1.5 \leq z \leq 4.5$, we would have expected
to find $10^{+3}_{-5}$ \civ\ doublets in our data, instead of only three,
adding to the evidence for a decrease in the frequency of intergalactic
\civ\ as we approach $z \sim 6$. 

4. Thus our data suggest a relatively rapid build-up of
intergalactic \civ\ over a period of only $\sim 300$\,Myr,
from  $\sim 0.9$ to $\sim 1.2$\,Gyr after the Big Bang.
Such a build-up could reflect the earlier rise of cosmic star formation 
activity from $z \sim 9$ indicated by galaxy counts and/or
an increasing level of ionisation in the IGM.
Numerical simulations of galaxy formation which include 
momentum-driven galactic outflows are
consistent with our findings and suggest that both factors---a
genuine build-up of metallicity and higher ionisation---contribute
to the rise in \Ociv\ from $z = 5.8$ to $z = 4.7$.

5. \emph{If} the value of \Ociv\ we have derived is representative
of the IGM at large, it corresponds to a metallicity
$Z_{\rm IGM} \geq  (9 \pm 5) \times 10^{-5}\,Z_{\odot}$.
The accumulation of this mass of metals in the IGM in turn
implies that only about one Lyman continuum photon per
baryon had been emitted by early-type stars
prior to $z = 5.8$. This background is clearly insufficient 
to keep the the Universe ionised, 
suggesting that more carbon is present at these early epochs,
in undetected ionisation stages, low column density systems, 
stars, or a combination of all three. 

6. All of these estimates are of necessity still rather uncertain.
However, with the forthcoming availability of X-shooter on the 
VLT,\footnote{See http://www.eso.org/sci/facilities/develop/instruments/xshooter/}
as well as planned improvements in the performance of near-IR spectrographs
on other large telescopes, it should be possible in the near future
to make full use of metal absorption lines in QSO spectra to trace
early star formation and the consequent reionisation of the IGM.

\section*{Acknowledgements}

It is a pleasure to acknowledge the expert 
assistance with the observations offered by
Rachel Gilmour at ESO, Grant Hill at Keck,
and Brad Holden at Santa Cruz. 
We are indebted to George Becker who generously provided
the efficient data reduction software used in the analysis of the
NIRSPEC spectra, and to Alice Shapley who very kindly helped
with its implementation and use in Cambridge. 
Jim Lewis also helped with the final processing of the data.
The paper was improved by valuable comments on
an early draft by George Becker, Claude-Andr\'{e} Faucher-Gigu\`{e}re, 
Martin Haehnelt, Ben Oppenheimer, and the referee Xiaohui Fan.
We thank the Hawaiian
people for the opportunity to observe from Mauna Kea;
without their hospitality, this work would not have been possible.
Some of this work was carried out during a visit
by Piero Madau to the Institute of Astronomy, Cambridge,
made possible by the Institute's visitor grant.
Support for this work was provided in part by 
NASA through grants HST-AR-11268.01-A1 and NNX08AV68G (P.M.).

\label{lastpage}

\end{document}